\documentclass[preprint]{aastex}
\graphicspath{{./}}

\date{\today}

\begin{document}
\title{Molecular Jet of IRAS 04166$+$2706}
\author{Liang-Yao Wang\altaffilmark{1,2,3}, 
            Hsien Shang\altaffilmark{2,3},
            Yu-Nung Su\altaffilmark{2}, 
            Joaqu{\'{\i}}n Santiago-Garc{\'{\i}}a\altaffilmark{4}, 
            Mario Tafalla\altaffilmark{5},          
            Qizhou Zhang\altaffilmark{6}, 
            Naomi Hirano\altaffilmark{2},
            \and  
            Chin-Fei Lee\altaffilmark{2}
            }
            
\altaffiltext{1}{Graduate Institute of Astronomy and Astrophysics, National Taiwan University, No. 1, Sec. 4, Roosevelt Road, Taipei 10617, Taiwan; lywang@asiaa.sinica.edu.tw}
\altaffiltext{2}{Academia Sinica, Institute of Astrophysics (ASIAA), P.O. Box 23–141, Taipei 106, Taiwan}
\altaffiltext{3}{Theoretical Institute for Advanced Research in Astrophysics (TIARA), Academia Sinica, P.O. Box 23-141, Taipei 10641, Taiwan}
\altaffiltext{4}{Instituto de Radioastronom{\'{\i}}a Milim{\'e}trica (IRAM), Avenida Divina Pastora 7, N{\'u}cleo Central, 18012 Granada, Spain}
\altaffiltext{5}{Observatorio Astron{\'o}mico Nacional (IGN), Alfonso XII 3, E-28014 Madrid, Spain}
\altaffiltext{6}{Harvard-Smithsonian Center for Astrophysics, 60 Garden Street, Cambridge, MA 02138, USA}

\begin{abstract}      
        The molecular outflow from IRAS 04166+2706 was mapped with the Submillimeter Array (SMA) at 350 GHz continuum and CO $J$ = 3--2 at an angular resolution of $\sim$1\arcsec. The field of view covers the central arc-minute, which contains the inner four pairs of knots of the molecular jet.  On the channel map, conical structures are clearly present in the low velocity range ($|$V$-$V$_0|<$10 km s$^{-1}$), and the highly collimated knots appear in the Extremely High Velocity range (EHV, 50$>|$V$-$V$_0|>$30 km s$^{-1}$). The higher angular resolution of $\sim$ 1\arcsec\ reveals the first blue-shifted knot (B1) that was missing in previous PdBI observation of \citet{santiago2009} at an offset of $\sim$6\arcsec\ to the North-East of the central source. This identification completes the symmetric sequence of knots in both the blue- and red-shifted lobes of the outflow. The innermost knots R1 and B1 have the highest velocities within the sequence. Although the general features appear to be similar to previous CO $J$ = 2--1 images in \citet{santiago2009}, the emission in CO $J$ = 3--2 almost always peaks further away from the central source than that of CO $J$ = 2--1 in the red-shifted lobe of the channel maps. This gives rise to a gradient in the line-ratio map of CO $J$ = 3--2/$J$ = 2--1 from head to tail within a knot. A large velocity gradient (LVG) analysis suggests that the differences may reflect a higher gas kinetic temperature at the head. We also explore possible constraints imposed by the non-detection of SiO $J$ = 8--7.

\end{abstract} 


\section{INTRODUCTION}
    The outflow driven by the low-luminosity (0.4 $L_\sun$) Class 0 source \citep{bontemps1996} IRAS 04166+2706 (hereafter I04166) is one of the few outflows with an extremely-high-velocity component. It is also the first such source reported in the nearby Taurus molecular cloud at 140 pc by \citet{tafalla2004}. In their study, distinct velocity regimes were identified with respect to the systemic velocity of 6.7 km\,s$^{-1}$ (relative to the LSR) derived from NH$_3$(1,1): the standard-high-velocity (SHV) component corresponds to velocity displacements of 2 to 10 km\,s$^{-1}$, the intermediate-high-velocity (IHV) component of 10 to 30 km\,s$^{-1}$, and the extremely-high-velocity (EHV) component of 30 to 50 km\,s$^{-1}$. The division is clearly demonstrated in their CO $J$ = 2--1 spectra, where emission in the EHV regime appears to be a prominent and distinct feature from the slower SHV component \citep[spectra of other molecular lines such as SO also show distinct velocity components; see][]{tafalla2010}.
  
   These different velocity components in the spectra were also found to have remarkably different spatial distributions. \citet{santiago2009} observed a series of well-collimated knots in the EHV regime lying along the outflow axis (position angle $\sim30\fdg4$), and a pair of limb-brightened conical cavity walls close to the systemic velocity. The measured opening angle of the shell is much wider than that of the collimated EHV gas, and there is no sign of momentum transfer from the jet to the ambient because the EHV knots keep an almost constant mean velocity. The velocity structure suggests that the shell can be understood as a shear flow of accelerated ambient gas moving along the cavity walls. \citet{santiago2009} argued that the SHV shell component is unlikely accelerated by the EHV jet.  An unseen wide-angle wind component may be required to explain the simultaneous presence of both components in the case of I04166 outflow. The authors pointed out that a unified wind model that naturally produces high-velocity well-collimated jet and low-velocity shell-like swept up materials may be suitable \citep{shang2006}. For simplicity in this work we refer to the two distinct features as the ``EHV knots (jet)'' and the ``low-velocity conical shells''.
      
Position--Velocity (PV) diagrams cut along the I04166 outflow axis show an intriguing velocity pattern. Within each EHV knot, there is a characteristic internal velocity gradient in which the gas closer to the source (the ``tail'') moves faster than the gas further away from the source (the ``head''). One may intuitively think of a knot as a collection of gas ejected from the central source at more or less the same time but with different intrinsic velocities. In this case the gas with higher velocity is supposed to move faster and becomes the head of the knot. The observed opposite velocity structure suggests a different kinematic picture. \citet{santiago2009} attributed this peculiar structure to the projection of lateral expansion of material from the internal working surfaces \citep{raga1990}. Due to its clean velocity structure, the I04166 outflow may help shed light on the ejecting history of a young stellar system as well as the debated launching mechanisms \citep{shu2000,konigl2000}.

In this work, we report new interferometric observations of the I04166 outflow in the 350 GHz continuum and the CO $J$ = 3--2 line with the Submillimeter Array (SMA)\footnote{The SMA is a joint project between the Smithsonian Astrophysical Observatory and the Academia Sinica Institute of Astronomy and Astrophysics, and is funded by the Smithsonian Institution and the Academia Sinica.}. We examine the four innermost pairs of knots at a higher angular resolution of $\sim1\arcsec$. Our results are complementary to the CO $J$ = 2--1 data from \citet{santiago2009}, and reveal more kinematic and spatial details that are valuable for potential models. This paper is organized as follows. Observations and data reduction are described in Section \ref{sec_obs}, and results are presented in Section \ref{sec_result}. Sections \ref{sec_discussion} and \ref{sec_conclusion} follow as discussions and conclusions.

\section{OBSERVATIONS AND DATA REDUCTION}\label{sec_obs}

        Observations toward I04166 were carried out with the SMA on 2010 September 9 and 22 in the extended configuration, and on 2010 October 5 in the compact configuration \citep[see][for information on SMA]{ho2004}. The total projected baselines range from 12 m to 225 m. All eight antenna were available on September 22, and seven were available on September 9 and October 5. We tuned the receivers to the SiO $J$ = 8--7 line at 347.331 GHz in the lower sideband so that the two sidebands, upper and lower, covered the frequency ranges from 355.6 to 359.6 GHz and 343.6 to 347.6 GHz, respectively. This setup allowed to observe simultaneously CO $J$ = 3--2 at 345.796 GHz and SiO $J$ = 8--7, although SiO $J$ = 8--7 was not detected. The spectral resolution for CO line was 0.8125 MHz and the corresponding velocity resolution was $\sim0.7$ km\,s$^{-1}$. The data were smoothed to 2 km\,s$^{-1}$ for further analysis. The primary beam size of the SMA antenna at this frequency is $\sim34$\arcsec, and three pointings separated by 18\arcsec\ along the outflow axis were observed to cover the four pairs of EHV knots closest to the central source. The phase center was set to [$\alpha(J2000)=$ 04$^{\rm h}$19$^{\rm m}$42$\fs$5, $\delta(J2000)=$ 27{\arcdeg}13{\arcmin}36{\farcs}0], which is the position of continuum peak determined in \cite{santiago2009}.

        The visibility data were calibrated using the MIR software package adapted for the SMA from the MMA software package developed originally for the Owens Valley Radio Observatory \citep{scoville1993}. We used 3C 84 as the gain calibrator, and Callisto as the flux calibrator. Uranus and 3C 454.3 were used as the passband calibrators for the compact configuration data, while only 3C 454.3 was used for the extended configuration data. The calibrated visibility data were imaged and then CLEANed with the MOSSDI task in the MIRIAD package \citep{sault1996} to produce CLEANed component maps. The final maps were obtained by restoring the CLEANed component maps with a synthesized (Gaussian) beam fitted to the main lobe of the dirty beam.
        
        The synthesized beam size of the CO channel maps was $1\farcs02\times0\farcs80$ at a position angle of $-84\fdg9$ with natural weighting. The resulting root-mean-squared (rms) noise level ($\sigma$) was $\sim37$ mJy\,beam$^{-1}$ in a 2 km\,s$^{-1}$ wide channel. The continuum map at 350 GHz ($\lambda=$ 850 $\micron$) was obtained by averaging all available line-free chunks of both sidebands using uniform weighting. The synthesized beam size of the map was $0\farcs85\times0\farcs59$ at a position angle of $-85\fdg1$. The rms noise level of the map is 2.8 mJy\,beam$^{-1}$.  The information is also summarized in Table \ref{tab_bsize_rms}.
        % rms measured in cleaned continuum map 
        % compact: no ant. 4 || extended1: no ant.6 || extended2: all ant. available         

\section{RESULTS}\label{sec_result}

\subsection{The 350 GHz Continuum}% (old 3.1)

\begin{figure}[ht]        
        \includegraphics[width=\textwidth]{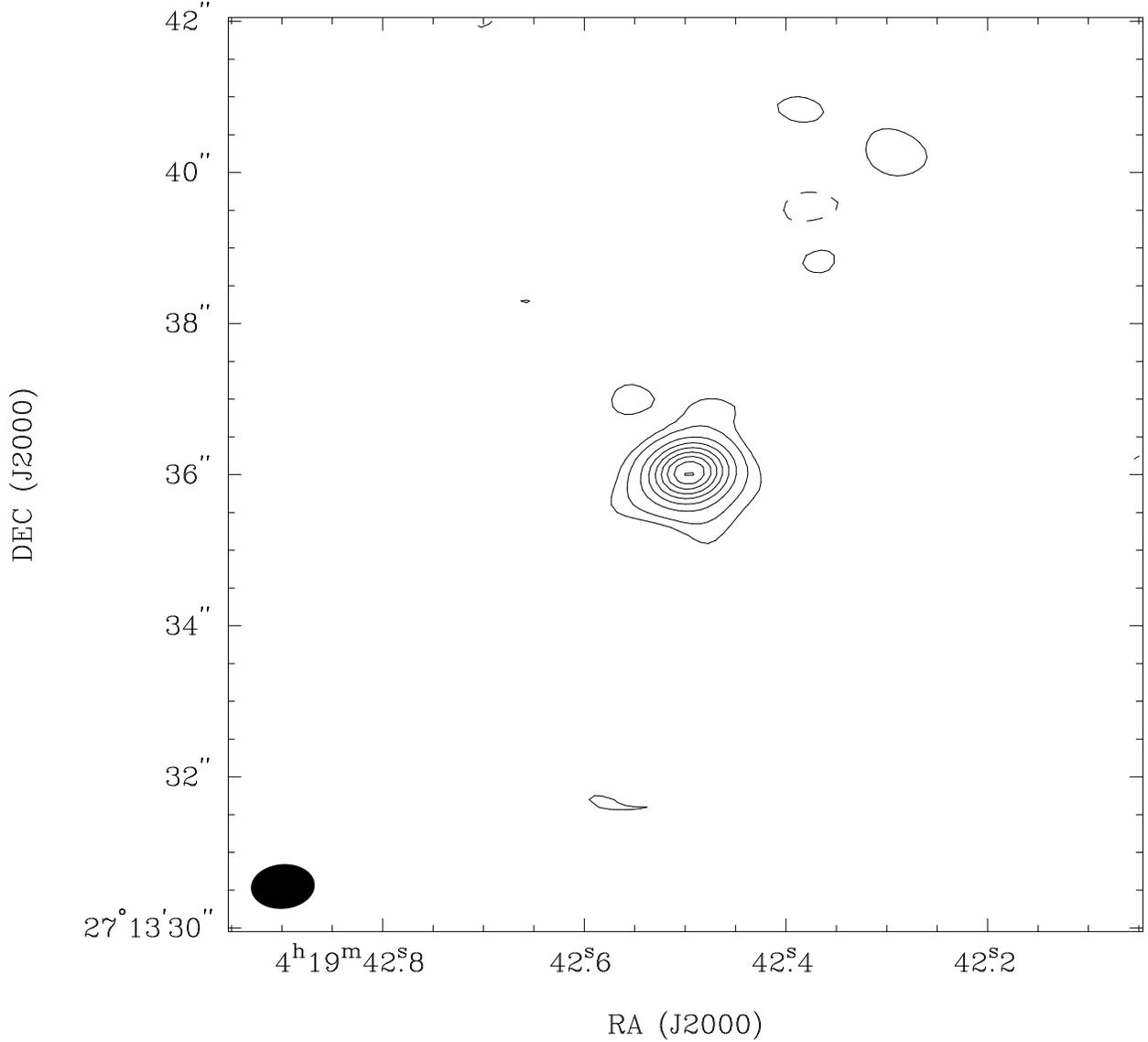}
        \caption{Continuum map of 350 GHz ($\lambda$ $850\micron$) emission toward the vicinity of I04166 made with uniform weighting. The map is made using all available visibility data in the line-free chunks. The source has a peak intensity of 135 mJy\,beam$^{-1}$. Contours are drawn at 3$\sigma$, 6$\sigma$, 12$\sigma$, 18$\sigma$, 24$\sigma$, 30$\sigma$, 36$\sigma$, and 42$\sigma$, where 1$\sigma$ is 2.8  mJy\,beam$^{-1}$. The FWHM of the synthesized beam is $0\farcs85\times0\farcs59$ at a position angle $-85\fdg1$.
        }
        \label{fig_continuum}
\end{figure}

    In Figure \ref{fig_continuum} we show the 350 GHz ($\sim850\micron$) continuum map toward the vicinity of I04166. Only one source with peak intensity 135 mJy\,beam$^{-1}$ was detected. Fitting the map with a single Gaussian component using MIRIAD task ``IMFIT'', we find the resulting center position to be [$\alpha(J2000)=$ 04$^{\rm h}$19$^{\rm m}$42{\fs}496, $\delta(J2000)=$ $+$27{\arcdeg}13{\arcmin}36{\farcs}013] with position errors 0\farcs018 and 0\farcs014 in $\alpha$ and $\delta$, respectively. This is nearly identical to the source position identified at 1.3 mm and 3.5 mm wavelengths and very close to the nominal position of IRAS 04166+2706 \citep{santiago2009}.
%IRAS04166 nominal position at $\alpha$($J$2000)=04$^{\rm h}$19$^{\rm m}$42{\fs}6, $\delta$ ($J$2000)= $+$27{\arcdeg}13{\arcmin}38{\arcsec}

\begin{figure}[ht]
        \includegraphics[angle=-90,width=\textwidth]{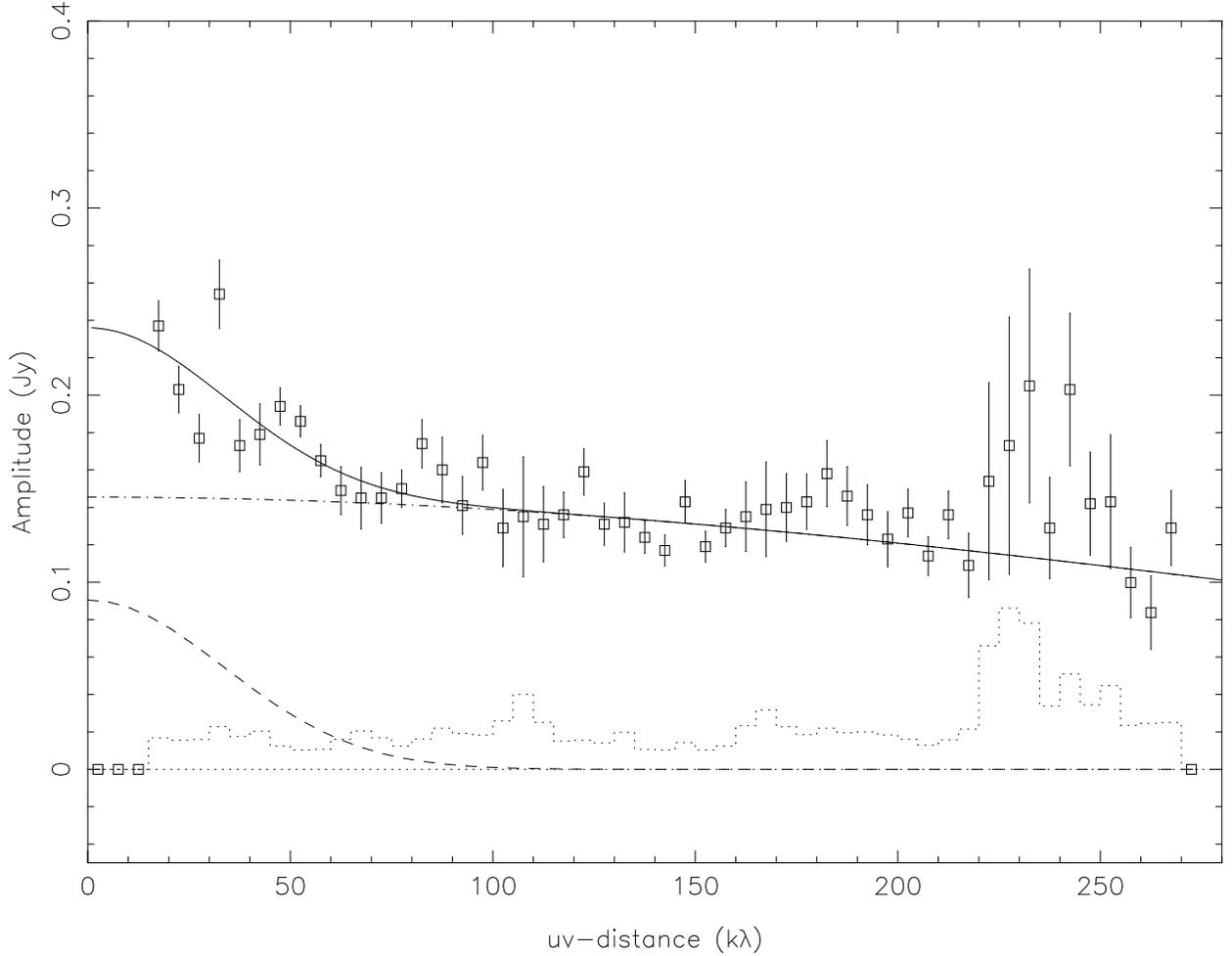}
        \caption{Plot of visibility amplitude vs. \textit{uv} distance for the 350 GHz continuum emission with 1$\sigma$ error bars. The dotted histogram is the expected amplitude for zero signal. We fit the profile with two Gaussian components. The dashed line is the curve for the extended component, and the dashed-dotted line is for the compact component. The solid curve is the total amplitude of the two components.
        The flux and FWHM of the two Gaussian components are 146 mJy, 0\farcs22 and 90 mJy, 2\farcs3 respectively.     
        }
        \label{fig_vis_fitting_2g}
\end{figure}

        In Figure \ref{fig_vis_fitting_2g} we examine the visibility amplitudes as a function of \textit{uv} distance. The visibility profile suggests that the continuum emission can not be described by a single Gaussian in the \textit{uv} domain. The source appears to consist of two components: an extended envelope that contributes to the flux at \textit{uv} distances $\lesssim 80$ k$\lambda$, and a compact source that is prominent at $> 80$ k$\lambda$. The latter is at most marginally resolved in our observations. To characterize the two components, we have fitted the visibility data with Gaussians using the MIRIAD task UVFIT. One Gaussian has a FWHM of $\sim2\farcs3$ and $90\pm15$ mJy integrated flux, and the other has a FWHM of $\sim0\farcs22$ and $146\pm6$ mJy integrated flux. The two components have therefore comparable flux, but their sizes differ by about one order of magnitude. The size scales correspond to $\sim$320 AU for the larger Gaussian and $\sim$30 AU for the smaller one, at a distance of 140 pc. Similar two-component structure, often interpreted as an envelope and a compact source, has also been observed in other Class 0 sources such as HH 211 \citep{lee2007b}, HH 212 \citep{codella2007,lee2007a}, and  L1448C \citep{schoeier2004,hirano2010}.

The frequency-dependence of the continuum flux could tell the dust properties of the source. We assumed a power-law frequency dependence of dust opacity \citep[ $\kappa_\nu \propto \nu^{\beta}$,][]{beckwith1990}, and compared the continuum flux at different wavelengths to estimate $\beta$. For simplicity, we also assumed an isothermal source and an optically thin emission so that the flux is simply proportional to $\nu^{2+\beta}$ in the Rayleigh--Jeans limit. The total flux at $850\micron$ is $\sim236$ mJy, and we compared this with previous measurements of 59 mJy and 11 mJy at 1.3 mm and 3.5 mm, respectively \citep{santiago2009}. Comparison with the former gives $\beta=1.28$, which is within the range of 1--2 often found for embedded YSOs \citep{dent1998}. On the other hand, comparison with the latter will result in a much smaller value of $\beta=0.17$. We note that in the above comparison the different \textit{uv} sampling at the three frequencies was not taken into account, and the physical scales probed may be different. The beamsize of 1.3 mm data is more comparable to our $850\micron$ data, while the 3.5 mm beamsize is much larger. This could explain the discrepancy in the derived $\beta$ value.

% Both fitting has rms residual of 3.128E-01
% compact component:  size 0.2174"+-0.0407" /  flux 145.9 +- 6.26 mJy 
% extended component: size 2.2599"+-0.318"   /  flux 90.2 +- 15 mJy 
% The fitting converge to rougly the same results for initial value of FWHM 1-6"
% reference of the 2-componet interpretation is from Lee2007 (HH211)

\subsection{Overall Morphology of the CO $J$ = 3--2 Molecular Outflow}% old 3.2

\begin{figure}[ht]
        \includegraphics[width=\textwidth]{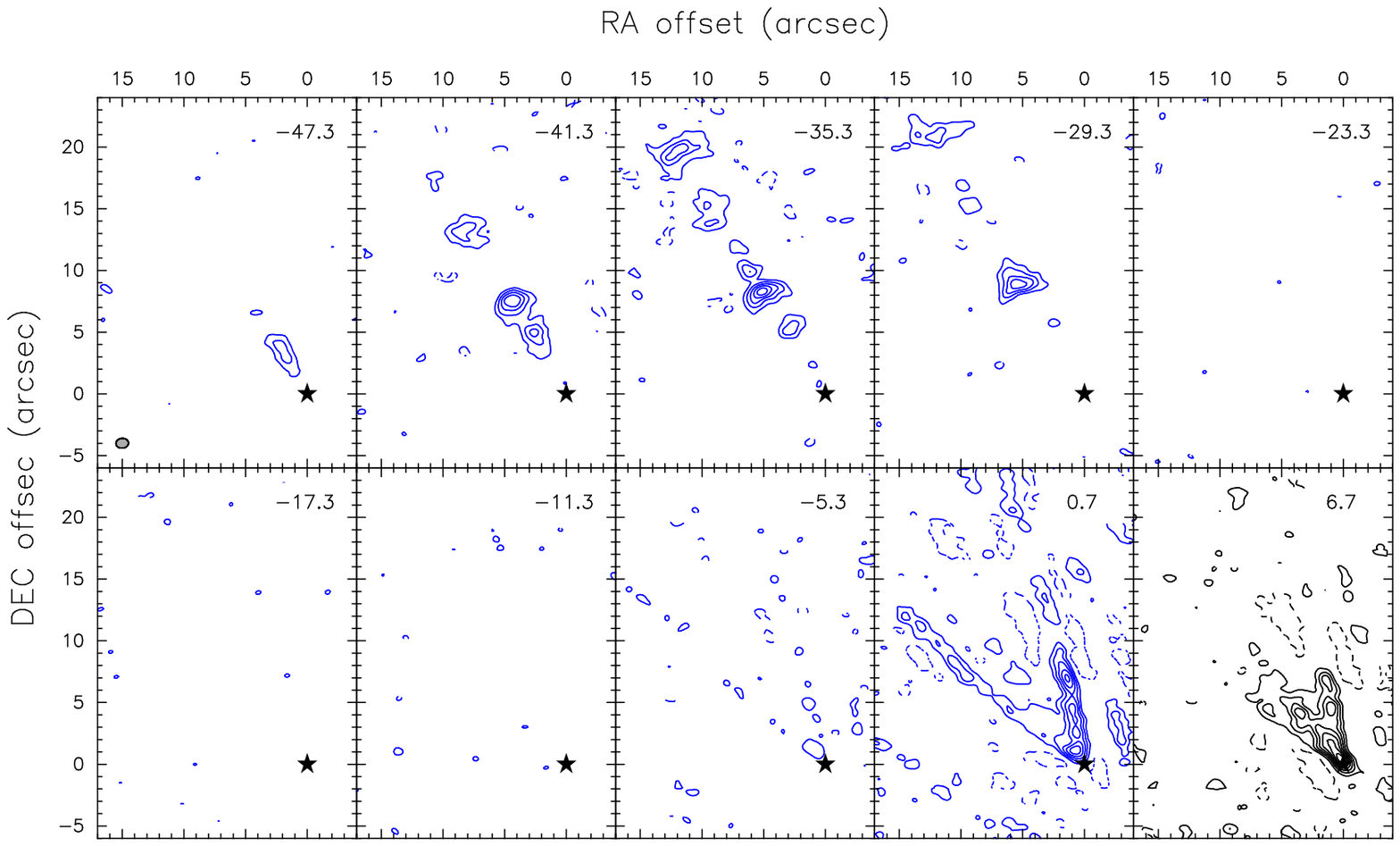}
        \includegraphics[width=\textwidth]{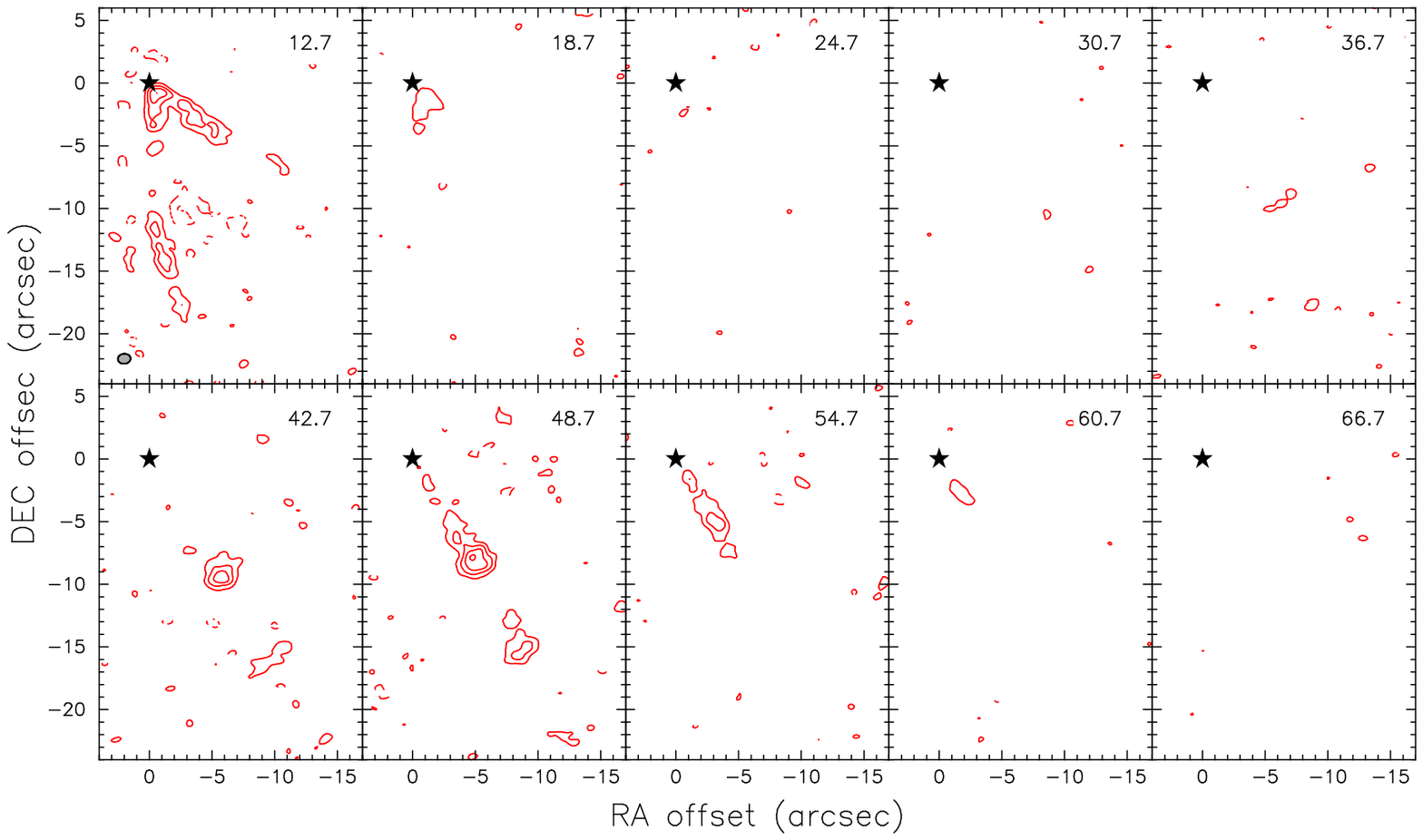}
        \caption{Channel maps of the CO $J$ = 3--2 line emission averaged over 6 km\,s$^{-1}$ velocity bin. The velocities (LSR) of the channels are shown in the upper right corner of each plot (system velocity is 6.7 km\,s$^{-1}$ ). The star symbol marks the position of the continuum source. Contour levels are from 4$\sigma$ in steps of 4$\sigma$, where 1 $\sigma$ is 21 mJy\,beam$^{-1}$. The FWHM of the synthesized beam  is $1\farcs02\times0\farcs80$ at a position angle $-84\fdg9$.      
         }
        \label{fig_co32_chmap_bin6}
\end{figure}

        The CO $J$ = 3--2 emission was detected in velocity ranges of $-58$ to $-34$, $-8$ to $+10$, and $+34$ to $+54$ km\,s$^{-1}$ with respect to the systemic velocity ($V_0=6.7$ km\,s$^{-1}$). For display purposes, in Figure \ref{fig_co32_chmap_bin6} we show the channel maps averaged over 6 km\,s$^{-1}$ wide velocity bins. The central LSR velocity is labelled on the top-right corner. 

    From top-left to bottom-right panels, channel maps are shown in the order of increasing radial velocities, i.e., from blue-shifted to red-shifted. In the first row, where the highest-speed blue-shifted emission ($V-V_0<-30$ km\,s$^{-1}$) is shown, one finds a series of emission peaks lying to the northeast of the central continuum source (the star symbol at the origin). Some of these knots appear to be compact, and some appear to be flattened and laterally elongated (e.g., the knots at $\sim9\arcsec$ and $\sim23\arcsec$ to the northeast of the central source in the $-35.3$ km\,s$^{-1}$ and $-29.3$ km\,s$^{-1}$ channels). All of the observed knots are well aligned in a straight line. In the second row, one finds no CO $J$ = 3--2 detections in the intermediate velocity channels until close to the systemic velocity. Emission in the $-6$ km\,s$^{-1}$ channel clearly delineates a V-shaped structure opening to the northeast direction with its apex at the position of the continuum source. Emission near $V_0$ (last panel of the second row)  appears to have a conical V-shaped structure to the northeast as well.
                
        The red-shifted emission shows remarkable symmetry with respect to the blue-shifted emission. The emission in the $+12.7$ km\,s$^{-1}$ channel (first panel of the third row) delineates a V-shaped structure opening to the southwest with a common apex as its northeast, blue-shifted counterpart. Like the blue-shifted lobe, there is no detection in the intermediate velocity range of the red-shifted emission. The highest-speed red-shifted emission is shown in the last row, and a series of emission peaks are clearly seen to the southwest of the continuum source. 
        
        To summarize, the bipolar outflow appears to be highly symmetrical with respect to the central continuum source. In both the blue- and red-shifted lobes, two distinct features appear at two different velocity regimes. Series of well-aligned knots are seen in the EHV range, while conical V-shaped structures are found in low-velocity channels.
        
\begin{figure}[ht]
        \includegraphics[width=0.8\textwidth]{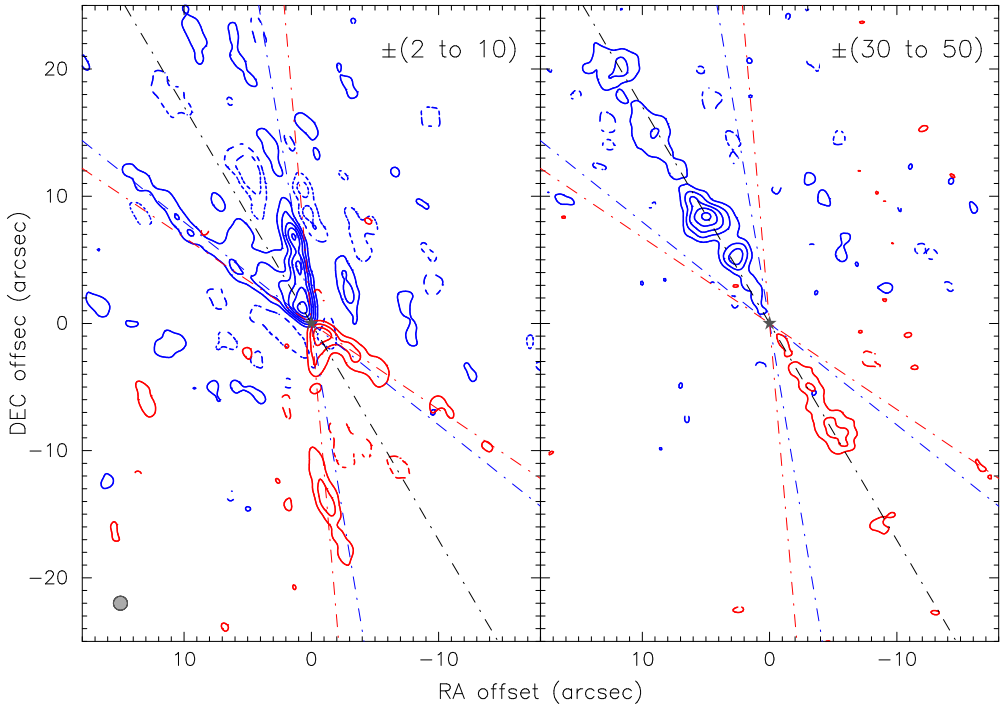}
        \caption{Integrated CO $J$ = 3--2  maps of the I04166 outflow over different velocity ranges demonstrating the low velocity V-shape conical shell (left panel, $10>|V-V_0|>2$ km\,s$^{-1}$) and the EHV jet (right panel, $50>|V-V_0|>30$ km\,s$^{-1}$). The data was first smoothed to a resolution of 1.1\arcsec\ before plotting. In the low velocity maps, contour levels are from 1.14 Jy\,beam$^{-1}$ km\,s$^{-1}$ in steps of 1.14 Jy\,beam$^{-1}$ km\,s$^{-1}$.
In the EHV map, contour levels are from 1.2 Jy\,beam$^{-1}$ km\,s$^{-1}$ in steps of 1.2 Jy\,beam$^{-1}$ km\,s$^{-1}$. The opening angles obtained for blue- and red-shifted CO $J$ = 3--2 low velocity shells (42{\arcdeg}/51{\arcdeg}) are illustrated with dot-dashed line of the corresponding color (blue/red).         
         }
        \label{fig_co32_mom0_SHVEHV}
\end{figure}

        % EHV flux sum: blue 4.059e3 / red 2.683e3 ratio 1.512
        To better illustrate the symmetry of the emission, we present in Figure \ref{fig_co32_mom0_SHVEHV} the integrated maps over the low-velocity ($10>|V-V_0|>2$ km\,s$^{-1}$) and EHV ($50>|V-V_0|>30$ km\,s$^{-1}$) ranges in the left and right panels, respectively. All data are smoothed to a beam size of 1\farcs1 before plotting. The integrated low-velocity maps clearly display the blue-shifted and red-shifted V-shaped structures. The two lobes do not overlap with each other, and they have a common apex at the position of the central source. In the integrated EHV maps, collimated blue- and red-shifted jets consisting of a series of knots are present. The position angle of the outflow axis was found to be 30\fdg4 in the previous CO $J$ = 2--1 study \citep{santiago2009}, and is shown with a dark dash-dotted line in the plot. The dash-dotted lines illustrate the opening angle of the low velocity shell. Detailed examination reveals more subtle structures which we discuss after consideration of the missing flux.

\subsection{The Missing Flux}% old 3.3

\begin{figure}[ht]
        \includegraphics[width=0.7\textwidth]{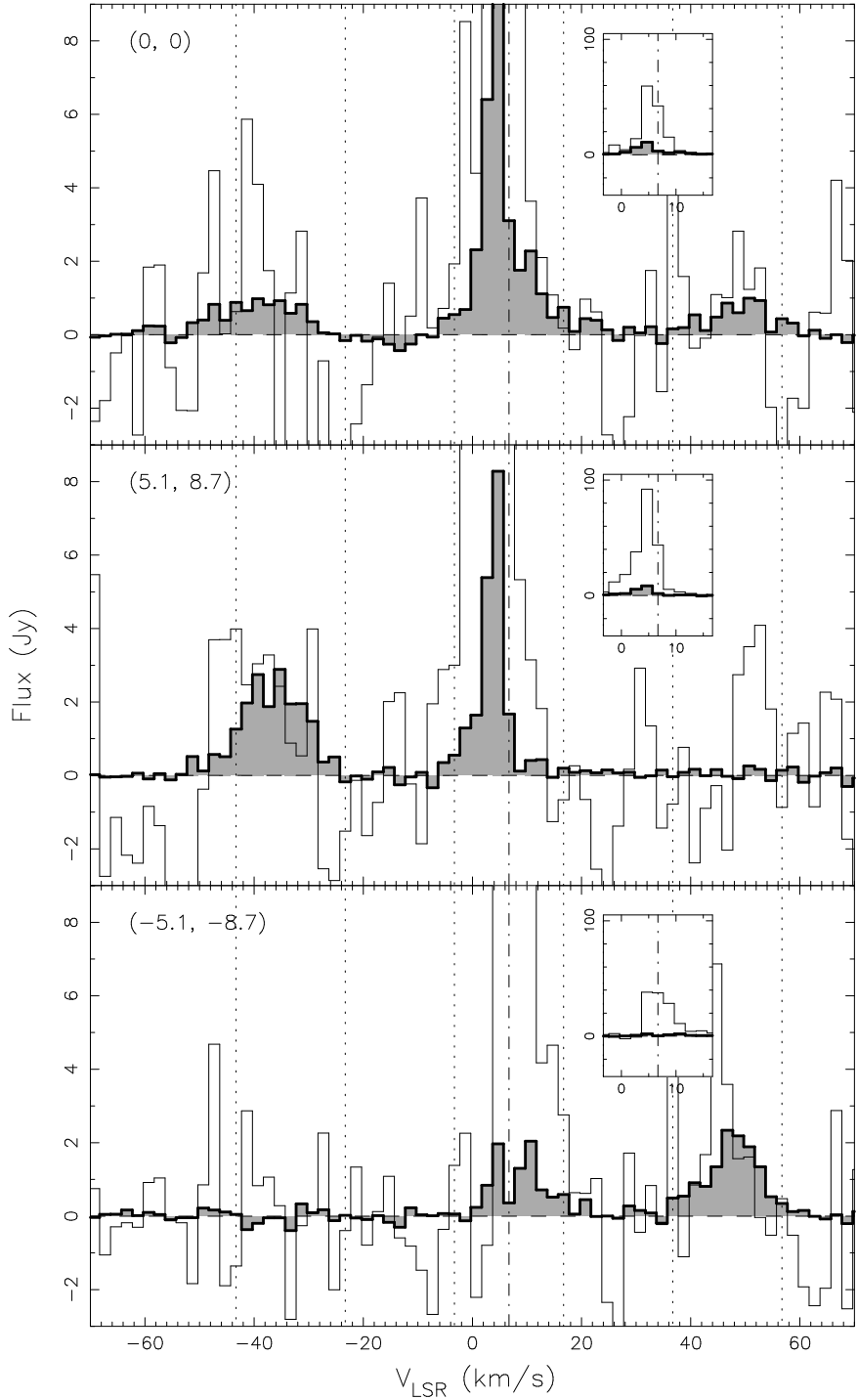}
        \caption{Spectra comparing the SMA interferometric CO $J$ = 3--2 data (gray histogram) to the JCMT single-dish data (white histogram) at three positions. The offsets from the continuum source of the positions are labelled in the upper left corner of each plot. The SMA data were convolved to a 16\arcsec\ beam in order to have the same spatial resolution as the JCMT data. The vertical dash-dotted lines show the systemic velocity of I04166 outflow (6.7 km\,s$^{-1}$), and the dotted lines illustrate the boundaries of velocity regimes at $\pm$10, 30, and 50 km\,s$^{-1}$ with respect to systemic velocity. A sub-plot inside each panel shows the low velocity part in a larger scale. 
         }
        \label{fig_co32_spec}
\end{figure}

        Interferometric observations inevitably suffer from the problem of missing flux due to their finite shortest baseline. Single-dish data can help determine how much flux is lost. The I04166 molecular outflow was covered in a James Clerk Maxwell Telescope (JCMT) Legacy Survey project whose data were smoothed with a Gaussian kernel that results in a final resolution comparable to a point source of $\sim 16\arcsec$ \citep{davis2010}. In Figure \ref{fig_co32_spec} we plot the spectra at the position of the central source and positions where the blue- and red-shifted emissions reach their peak values on the convolved SMA map. The resulting spectra show the SMA data recover only about 6\% to 20\% of the flux within the low-velocity regime ($|V-V_0|<10$ km\,s$^{-1}$), depending on the positions. On the other hand, there is no clear detection within the EHV regime in the single-dish spectrum due to its higher noise level. This implies a significantly lower fraction of the missing flux in the EHV regime.

\subsection{The EHV Jet}
        
\subsubsection{Jet Kinematics along the Axis}\label{subsec_jet_kinematics_along_the_axis}   

        Our observations cover the four inner pairs of knots of the EHV jet at $\sim1\arcsec$ resolution. We overlay in Figure \ref{fig_co32co21_pv} a PV diagram of our CO $J$ = 3--2 data (red contours) on an equivalent diagram of the CO $J$ = 2--1 data from \citet{santiago2009} (green contours). The blue-shifted emission to the northeast of the continuum source appears on the lower-left quarter of this diagram, and the red-shifted, southwest emission appears on the upper-right. As already seen in the channel maps, the CO $J$ = 3--2 emission is only found within the low-velocity regime ($|V-V_0|<10$ km\,s$^{-1}$) and the EHV regime ($|V-V_0|>30$ km\,s$^{-1}$). While the symmetrical morphology of the bipolar outflow is clearly seen in the integrated map (Figure \ref{fig_co32_mom0_SHVEHV}), the PV diagram in Figure \ref{fig_co32co21_pv} further demonstrates the symmetry in kinematics. Both blue- and red-shifted EHV gas present a sawtooth-like pattern in which the magnitude of velocity varies semi-periodically with offsets from the central source. Each emission peak has a maximum speed at its ``tail'' (closer to the driving source) and decreases outwards toward its ``head''. The vertical, solid line segments near the bottom of the PV diagram label the positions of the the EHV knots determined in \citet{santiago2009}. They performed two-dimensional Gaussian fitting to the integrated CO $J$ = 2--1 emission over the entire EHV regime and identified the positions for all the knots but B1 which was unclear in their data. 
  
  	While the sawtooth velocity pattern seen in CO $J$ = 3--2 resembles that of CO $J$ = 2--1, our new data reveal a number of new features. First, the higher angular resolution allows us to clearly distinguish the innermost blue-shifted knot B1 at an offset of $\sim6\arcsec$ to the northeast of the central source. The presence of B1 was previously missed due to its proximity to B2. By confirming this counterpart to the first red-shifted knot R1, the symmetry of the jet is now more completely mapped. Furthermore, the observed maximum speeds within knots B1 and R1 are $\sim5$ km\,s$^{-1}$ greater than B2, R2, and other more distant knots. Finally, as the PV diagram shows, some emission is detected between B2 and B3 which makes the boundary blurry. For clarity, we find it desirable to define the domain (section) of each knot in addition to their peak position. Taking the semi-periodic behaviour into account, the jet can be naturally divided into individual sections using the positions of ``slow heads". For example, the emission in region between the slow head of B2 (at $\sim11\arcsec$ offset) and that of B1 (at $\sim6\farcs5$ offset) can be regarded as knot B2. In Figure \ref{fig_co32co21_pv} we label the sections with thick horizontal bars on top and their boundaries with vertical dotted lines.                                                                        
                                     
          In Figure \ref{fig_co32co21_pv_and_ehv_5kms_chmap_hori} a closer view of the EHV knots is shown.  On the top, a PV diagram of the EHV emission is plotted for both lobes with absolute velocity increasing upwards. For direct reference we show the maps of the corresponding velocity regime below with emission averaged over 5 km\,s$^{-1}$ velocity bins. From panels (a) to (f), channel maps are plotted in a sequence of decreasing speed offset from $V_0$, and are all rotated counterclockwise by 59\fdg6 so that the northeast lobe of the outflow points to the left and the southwest lobe to the right. The moderate bin width enhances the signal-to-noise ratios while preserving velocity information we are interested in. Here we first focus on the sawtooth velocity pattern on the top panel.
                   
          The velocity pattern is better resolved in our high angular resolution data, which allows a quantitative inspection. . We characterize the internal velocity gradient of a knot by measuring its ``slope'' on the PV diagram. The number describes how rapidly the (radial) velocity of the gas drops with (projected) distance from the central source, and could be potentially useful for future dynamical analysis. To do this, we first perform two-dimensional Gaussian fitting in 2 km\,s$^{-1}$ bin channel maps to determine the knot positions. A slope is then determined by fitting a straight line to the points. The positions are marked with solid triangles on the PV diagram of Figure \ref{fig_co32co21_pv_and_ehv_5kms_chmap_hori} and the fitted lines are also shown. We note that in some cases the knot emission is too fragmented for positions to be reasonably determined and is therefore ignored in this analysis. The resulting velocity gradients range from $\sim2.3$ to $\sim6.4$ km\,s$^{-1}$\,arcsec$^{-1}$, or  $\sim3.4\times10^3$ to $\sim9.4\times10^3$ km\,s$^{-1}$\,pc$^{-1}$. It appears that in general the slope of a blue knot is steeper than that of its red counterpart, with B3/R3 being the exception. Although the sampling space is small, there is also a possible trend for the more distant knots to have shallower gradients. Since a larger distance from the source is equivalent to a longer dynamical age, the correlation between slopes and distances may suggest that slopes can evolve over time. Knot R1 appears to be an outlier in this scenario, but it should be noted that there is a slight hint from the PV diagram that knot R1 (and also B1) may consist of two or more sub-knots (sub-sections). Data with better sensitivity and angular resolution are needed to draw a definite conclusion.

\subsubsection{Position Offsets between the CO $J$ = 3--2 and CO $J$ = 2--1 Peaks}\label{subsec_co_peak_offset} % old 3.4.3

        A Comparison between the CO $J$ = 3--2 and CO $J$ = 2--1 emissions shows interesting results. Despite the general agreement of a sawtooth pattern in the PV diagram (Figure \ref{fig_co32co21_pv}), the peaks of CO $J$ = 3--2 do not completely coincide with those of CO $J$ = 2--1. Instead, at a given velocity, the CO $J$ = 3--2 emission tends to peak further toward the head of each knot. This phenomenon is clearly seen among the red-shifted EHV knots as well as the blue-shifted knot B1, but is unclear or ambiguous for knots B2, B3, and B4. We can take a closer look at the peak positions in the channel maps of Figure \ref{fig_co32co21_pv_and_ehv_5kms_chmap_hori}. One can clearly see displaced peak positions of CO $J$ = 3--2 and CO $J$ = 2--1 in R2 and R3 in panels (d), (e) and (f). The former tends to peak further away from the central source than the latter. Knots R4 and B4 in panels (d) and (e) also show a hint of the same phenomenon in spite of much weaker detections. To estimate the offsets quantitatively, we fit two-dimensional Gaussians to the CO $J$ = 3--2 and CO $J$ = 2--1 data and obtain their peak positions. Differences between their distances from the central source are calculated. The dark and gray triangles on the channel maps in Figure \ref{fig_co32co21_pv_and_ehv_5kms_chmap_hori} label the fitted positions of CO $J$ = 3--2 and CO $J$ = 2--1 knots, respectively. We find an average offset of 1\farcs42 for the red-shifted knots and only 0\farcs15 for the blue-shifted ones. This confirms our observation that an offset of peak positions is present in the red-shifted lobe. We defer the discussion of possible uncertainties to section \ref{subsec_discussion_peak_offset_errors}.
        
Assuming that both the CO $J$ = 3--2 and CO $J$ = 2--1 emissions come form the same bulk CO gas, there are two possible interpretations of the offsets found in the PV diagram. One is that gas found in a given velocity bin is systematically brighter in CO $J$ = 3--2 further away from the central source on the channel maps. Alternatively, at a given position, the higher velocity gas systematically has stronger CO $J$ = 3--2 emission. This latter interpretation is well shown in Figure \ref{fig_co32co21_pv_crosscut}, where the CO $J$ = 3--2 emission tends to peak at a higher velocity than CO $J$ = 2--1 at a given position. The current data do not seem to distinguish the two interpretations. Some degree of uncertainty remains when one directly compares the peak positions in the two sets of data, and the phenomenon should be examined more carefully. The phenomenon is likely real, however. The systematically different peak positions between CO $J$ = 3--2 and CO $J$ = 2--1 could be an indicator of different excitation conditions within each knot. A trace structure in temperature and/or density could be present, as is expected in the presence of an internal shock.

\subsubsection{Head--Tail Structure of the EHV Peaks} % old 3.4.2

We examine the structures of knots in details, and the linkage between their morphology and kinematics is revealed. In Figure \ref{fig_co32co21_pv_and_ehv_5kms_chmap_hori} the divisions of knots can be seen with vertical dotted lines across the PV diagram and the channel maps, and the knot morphology in different velocity bins can be examined. For example, knot B2 is well detected and a gradual change in the shape of CO $J$ = 3--2 emission can be seen across the velocity range of 35--55 km\,s$^{-1}$ (panels b to e). In panel (b), the highest velocity gas appears as a small peak slightly elongated along the outflow axis. The emission is oval-shaped in panel (c) and is slightly elongated perpendicular to the jet axis. In panels (d) and (e), the emission appears cap-like and is more flattened. In short, the lower velocity gas of the knot in panels (c), (d), and (e), is more laterally broadened. A similar change of morphology across the velocity bins is also found in knot R2. Emission from other knots is weaker, especially in the higher velocity channels, but in panels (d) and (e) we can find similar flattened structures in knots R3, R4, and B4. The emission at the head of each knot tends to have a more flattened structure and appears broader than that of the tail. 
                  
The structure of the EHV knots can be further examined in Figure \ref{fig_co32co21_pv_crosscut} with a series of position--velocity diagrams cut perpendicular to the jet axis at different offsets from the central source. For both the blue-shifted (upper panels) and red-shifted lobes (lower panels), offsets ranging from $\pm$ (7\arcsec\ -- 11\arcsec) are plotted to cover knots B2 and R2, which display the clearest structures. The CO $J$ = 2--1 data are plotted with green contours for comparison. The diagram is plotted in a way that the jet velocity increases upward in both blue- and red-shifted lobes. Zero position offset corresponds to position of the jet, and increasing offsets (toward the left) correspond to increasing spatial offsets toward southeast direction. The two panels farthest to the left ($\pm7\arcsec$) show the tails of knots B2 and R2, and those farthest to the right ($\pm11\arcsec$) correspond to their heads.  The gradual change in the lateral size of a knot with velocity can be seen in this diagram by comparing the width of the emission in different velocity slices. The head-tail structure is visible if we compare the higher-speed narrow emission in panels $\pm07\arcsec$ and $\pm08\arcsec$ to the lower-speed broader emission in panels $\pm10\arcsec$ and $\pm11\arcsec$. A trend of decreasing jet velocity across the panels from left to right simply reflects the velocity gradients along the jet axis within knots B2 and R2.

\subsubsection{CO $J$ = 3--2/CO $J$ = 2--1 Line Ratios in EHV Knots} % old 3.4.4

The excitation condition of the EHV gas is best illustrated using the CO $J$ = 3--2/$J$ = 2--1 line-ratio maps in Figure \ref{fig_co32co21_pv_and_ehv_5kms_chmap_hori_ratio}. These line ratio maps were made from data binned by 5 km\,s$^{-1}$ with flux greater than 3$\sigma$, after being smoothed to the same spatial resolution of 3\arcsec. The spatial extent of the ratio map is primarily limited by the CO $J$ = 3--2 emission. The line-ratio PV diagram at the top is computed from data binned by 2 km\,s$^{-1}$ where only pixels with flux greater than 2$\sigma$ are adopted. The color scale ranges from 0 to 2 and the contours are drawn at a 0.3 interval. The ratio maps show that the gas with a higher line ratio ($>1.5$) mostly appears in the knots closer to the central source (B1, B2, R1, and R2), and at relatively higher velocities $|V-V_0|>45$ km\,s$^{-1}$, i.e., in panels (a), (b), and (c). The innermost knots R1 and B1 appear to have the highest CO $J$ = 3--2/CO $J$ = 2--1 line ratios among the knots. On the other hand, within the individual knots, higher line ratios are found close to the head of each knot. This is especially true for R2 and R3 in PV and panels (d), (e), and (f). The PV diagram of ratios also displays the same phenomenon. This result is consistent with the finding in Section \ref{subsec_co_peak_offset} that the CO $J$ = 3--2 emission tends to peak toward the head compared to CO $J$ = 2--1. It shows that the offset is still visible after smoothing the data to the same spatial resolution of 3\arcsec.                         
    
\subsection{The Low-Velocity Conical Shells}\label{subsec_LVShell} % old 3.5

In the low-velocity regime, the CO $J$ = 3--2 emission clearly delineates two conical (V-shaped) shells opening to the northeast and southwest. The blue-shifted and red-shifted conical structures appear to be symmetrical and share a common apex at the position of the continuum source. The opening angle of the shells can be obtained by measuring the offset position of shell emission from the central source, and taking arctangent of the ratio between its perpendicular and parallel component with respect to the outflow axis. The average opening angles are thus found to be $\sim42\arcdeg$ for the blue-shifted lobe and $\sim51\arcdeg$ for the red-shifted lobe and are shown in Figure \ref{fig_co32_mom0_SHVEHV} with dash-dotted lines. We note that, as the figure shows, the red-shifted shell emission appears weaker and less continuous so its angle is relatively uncertain.

In Figure \ref{fig_co32co21_mom0_compare_SHV_opening} we overlay the low velocity shell of CO $J$ = 3--2 (blue and red contours) over that of CO $J$ = 2--1 from \citet{santiago2009} (shown in color scale). The maps are presented at a larger scale to provide a global comparison, and the white dashed circles are the fields of view of our observations. The left and right panels show the blue- and red-shifted lobes respectively, and one can see that the CO $J$ = 3--2 emission coincides with the CO $J$ = 2--1 data well in both cases. With a higher angular resolution, the CO $J$ = 3--2 data sharply traces the edges of the conical structures. The opening angles we obtained from the blue- (42{\arcdeg}) and the red-shifted shell (51{\arcdeg}) are illustrated with blue and red dash-dotted lines along the shells.  Note that although the values are different from the 32\arcdeg\ found in \citet{santiago2009} (illustrated with green dash-dotted lines here), Figure \ref{fig_co32co21_mom0_compare_SHV_opening} suggests that the discrepancy is due to the different outflow scales considered. The red and blue dash-dotted lines well approximate edges of the V-shape structure up to a distance of $\sim15\arcsec$ from the central source. But beyond this, the detected  CO $J$ = 2--1 emission is slightly curved toward the outflow axis and is better described by the green dash-dotted lines in the blue-shifted lobe. It therefore appears that the opening angle of the V-shaped structure is decreased and may be more parabolic- or hyperbolic-like on a larger scale.
     
The velocity structure of the conical V-shaped component across the outflow axis can be seen in the PV diagrams in Figure \ref{fig_co32co21_pv_crosscut}. Focusing on the low-velocity emission ($|V-V_0|<10$ km\,s$^{-1}$), the CO $J$ = 3--2 emission traces the edge of the CO $J$ = 2--1 emission like walls that separate the outflow from its ambient cloud. The position of such emission does not have any apparent velocity dependence across the range from 0 to $\pm8$ km\,s$^{-1}$. As mentioned in \citet{santiago2009}, this kinematic feature suggests that the outflow motions are directed mostly along the shells with negligible perpendicular velocity component \citep{meyers1991}, and can be naturally understood if the observed emission represents a shear flow of accelerated ambient gas moving along the wall of evacuated cavities on both sides.
        
       A larger opening angle can provide a stricter constraint on the inclination angle of outflow axis from the plane of sky. Given that no spatial overlap between the blue- and red-shifted low velocity shells is observed and that the velocity is likely directed along the shell, we expect the inclination to be greater than the half-opening angle of the cone in order to avoid overlaps between the lobes. Similarly, the axis should also differ from the line-of-sight direction by more than the half-opening angle of the cone. Taking an average opening angle of blue (42{\arcdeg}) and red (51{\arcdeg}) SHV shells to be $\sim$46\fdg5, the cone half-opening angle of $\sim23\arcdeg$ implies that the inclination angle of I04166 should lie between 23\arcdeg--67\arcdeg\ .

\section{DISCUSSION}\label{sec_discussion}

\subsection{Large Velocity Gradient Analysis of EHV Jet}\label{subsec_lvg_ehv}

We carry out large velocity gradient (LVG) calculation \citep{goldreich1974,surdej1977} and compare the results to our observations to constrain the physical properties of the jet \citep{nisini2007}. Since the LVG model considers the Doppler shift associated with the gas motions in the system to simplify the calculation, the velocity gradient of the gas is naturally a key parameter to the model. For the I04166 jet, we estimate the velocity gradient from the sawtooth velocity pattern on the PV diagram, and a value of $\sim5\times10^{3}$ km\,s$^{-1}$\,pc$^{-1}$ is adopted (see Section \ref{subsec_jet_kinematics_along_the_axis}). The collisional rate coefficients and energy levels are required for solving statistical equilibrium equations in the LVG model, and we obtained the data from LAMDA\footnote{the Leiden Atomic and Molecular Database, http://home.strw.leidenuniv.nl/$\sim$moldata/}. The datafile for CO is compiled from \citet{flower2001} and \citet{wernli2006}, and those for SiO from \citet{dayou2006}. Both of them include extrapolation to energy levels up to  $J$ = 40 and temperatures up to 2000 K following \citet{schoeier2005}. Finally with a chosen fractional abundance, we are able to calculate the line intensities and their ratios over a range of  (H$_2$)  density (10$^{3}$--10$^{10}$ cm$^{-3}$) and temperature (10-2000 K) . 

In general, comparison to the observed line ratio is preferred over the absolute flux of transitions. This is because the absolute flux is more vulnerable to the beam filling factor problem if there are small and clumpy structures within a beam. Since only one line ratio is available to us, we will also examine the result including the flux as a constraint. we discuss our LVG analysis using CO emission and explore possible constraints imposed by the non-detection of SiO $J$ = 8--7. It is important to note that the results are sensitive to the adopted velocity gradient and the value of fractional abundance, and the effect of varying these inputs will also be discussed.

\subsubsection{CO $J$ = 3--2 / CO $J$ = 2--1 Line Ratio}
We use the LVG model to calculate CO $J$ = 3--2/$J$ = 2--1 line ratios. A canonical value of 10$^{-4}$ is assumed for the fractional abundance of CO relative to molecular hydrogen. The results are in the left panel of Figure \ref{fig_i04166_lvg} labelled with dotted contours. To explore the conditions of the I04166 jet, we examine the more isolated knots R2 and R3 and take typical values of line ratio at the head and tail of the knots of $\sim$1.0 and 0.6. Contours of the two specific values are highlighted with thick dark lines in Figure \ref{fig_i04166_lvg}, and their loci on the temperature--density plane identify possible conditions capable of reproducing the observed flux and ratio. As illustrated, the observed line ratios could occur under a wide range of density--temperature conditions, and the line ratio alone places little constraint. If we also include the absolute flux of the transitions, as shown with red and blue contours, the density is found to be around $\sim8\times10^{3}$ cm$^{-3}$, and the kinetic temperatures around $\sim30-60$ K.  Furthermore, it appears that the LVG model predicts a higher line ratio for a higher temperature gas in general, so the observed line ratio may suggest a higher gas temperature toward the head. In principal, such a trend should be more robust than the constrained absolute scale of temperature or density, which is likely effected by the problem of beam filling factor and an additional error of $\sim20$\% expected in flux calibration.

\subsubsection{SiO $J$ = 2--1 And The Non-Detection of SiO $J$ = 8--7 Emission} % old 4.2.2

We also carried out LVG calculation of SiO $J$ = 8--7/$J$ = 2--1 line ratio. The results are in the right panel of Figure \ref{fig_i04166_lvg} (dotted contours). It should be noted that the fractional abundance of SiO relative to H$_2$ varies greatly under different conditions. We adopt $10^{-7}$ as a reasonable value for the case of I04166 jet, which was derived in \citet{tafalla2010} through analysis of population diagram under the local thermal equilibrium (LTE) assumption.

No SiO $J$ = 8--7 emission was detected in our SMA observations through the velocity ranges covered, but we may use the upper limit to constrain the physical conditions of the jet. The rms noise level of the SiO map, binned every 2 km\,s$^{-1}$, is 37 mJy per $1\farcs01\times0\farcs80$ beam, and we smoothed the data to an angular resolution of $4\farcs80\times3\farcs36$ in order to compare with the PdBI SiO $J$ = 2--1 data. The resulting $\sigma$ is 107 mJy\,beam$^{-1}$, or equivalently, 0.067 K, and we adopt 0.2 K (3$\sigma$) as the upper limit of SiO $J$ = 8--7 emission. On the other hand, the SiO $J$ = 2--1 peak emission is 1.3 K for the blue-shifted jet, which gives an SiO $J$ = 8--7/$J$ = 2--1 line ratio upper limit of  0.15; the red-shifted jet, with a 1.0 K SiO $J$ = 2--1 peak emission, gives an upper limit of 0.2 for the same line ratio. 
The thick dark lines in Figure \ref{fig_i04166_lvg} highlight these particular contour levels, and the loci are upper limits to jet conditions in a form of density--temperature pairs. For example, the gas temperature should not be higher than $\sim$500 K if the density is $\sim$10$^5$ cm$^{-3}$, and no higher than $\sim$100 K if the density is $\sim5\times10^5$ cm$^{-3}$. If we also include the observed SiO $J$ = 2--1 flux (blue and red lines) and the upper limit of SiO $J$ = 8--7 flux (green line) as constraints, the LVG model suggests an upper limit of $\sim700$ K for the kinetic temperature of the gas, and a possible density range of 3$\times$10$^4$ to 10$^5$ cm$^{-3}$ depending on the temperature. We note that this estimation of the density range is based on the absolute flux of SiO $J$ = 2--1, which may suffer from the beam dilution problem. The density appears lower than the typical value of 10$^6-$10$^7$ cm$^{-3}$ found in HH 211 through the SiO $J$=8--7 transition \citep{nisini2002,palau2006,hirano2006,lee2007b}.

\subsubsection{Effects of Varied Input Parameters}
       
Input parameters of the LVG model can greatly affect the results. We vary the input parameters and perform the same analyses to demonstrate the dependence of the constrained values on these chosen conditions. In the left two panels of Figure \ref{fig_i04166_lvg_uncertainty} we show the same CO calculation with a factor of two higher (10$^{4}$) and lower (2.5$\times10^{3}$) values of the velocity gradient. While the profile of CO $J$ = 3 --2/$J$ = 2--1 line ratio does not change significantly, we found that the absolute flux is rather sensitive to the inputs. Increasing the velocity gradient by a factor of two would give about a factor of two higher constrained density with a decreased kinetic temperature. On the other hand, in the right two panels of Figure \ref{fig_i04166_lvg_uncertainty} are the results of LVG calculation for SiO emission with different adopted fractional abundances. It turns out that if the fractional abundance is an order of magnitude lower (10$^{-8}$), the constrained upper limit of gas temperature will become much lower ($\sim100$ K), but the density range will be about an order of magnitude higher ($\sim4\times10^{5}$ cm$^{-3}$). If a ten times higher fractional abundance (10$^{-6}$) is adopted, we actually can not derive an upper limit of the temperature from the upper limit of line ratio. These comparisons suggests that the constrained absolute temperature and density scale (or an upper limit) using the absolute flux are rather sensitive to the adopted parameters, and the trend of relatively higher gas temperature at the head is comparatively robust.

\subsection{Uncertainties in The Analysis of Offsets in The Peak Positions}\label{subsec_discussion_peak_offset_errors} % old 4.1

 The displacement between the peak positions of the CO $J$ = 3--2 and CO $J$ = 2--1 EHV gas could be a clue to help uncover the process of knot formation. We discuss the uncertainties involved here for completeness. We did not include every knot in each channel (only those marked with solid triangles were included) when estimating the average peak offsets between the two rotational transition lines in Figure  \ref{fig_co32co21_pv_and_ehv_5kms_chmap_hori}, as in some cases a fair comparison is not possible due to the poorer resolution of the CO $J$ = 2--1 data. Taking knot B2 in panel (e) as an example, it would first appear that the CO $J$ = 2--1 emission around $\sim11\farcs5$ lies farther from the central source than its CO $J$ = 3--2 counterpart. At the same time, one would also notice that the patch of CO $J$ = 3--2 emission actually consists of two parts: a laterally-broadened structure which we recognize as the head of knot B2, and a more compact blob near $\sim12\arcsec$ lying just ahead of it. The two features are clearly separated in panel (d) and in the PV diagram. Given the proximity, the two components will be indistinguishable at 3\arcsec\ resolution, and only a single elongated feature is seen in the CO $J$ = 2--1 maps. Judging from the PV diagram, the blending is expected within the velocity range of $-37$ km\,s$^{-1}$ to $-43$ km\,s$^{-1}$, and panels (d) and (e) may be ignored and the peaks of knot B2 are compared in panel (c) to find that CO $J$ = 3--2 emission lies slightly farther than CO $J$ = 2--1 does. Unfortunately, the proximity of knot B1 and B2 in panel (c), by a similar argument, suggests that a direct comparison of peak positions is not possible. In other words, the incompatible spatial resolutions in the two data sets cause obvious problems in some cases that were excluded in our estimation of peak offsets. We note that the offset persists even when we smoothed the CO $J$ = 3--2 data to the same spatial resolution of the CO $J$ = 2--1 emission (3\arcsec). This is indirectly shown in the CO $J$ = 3--2/CO $J$ = 2--1 ratio map (Figure \ref{fig_co32co21_pv_and_ehv_5kms_chmap_hori_ratio}). That the higher line ratios are found toward the heads of the red-shifted knots R2 and R3 is equivalent to a stronger CO $J$ = 3--2 emission concentrated toward the heads.

It is also possible that the displaced peak positions are the consequence of proper motions between the two observations, i.e., 6 years between 2004 and 2010. Taking 140 pc to be the distance of Taurus molecular cloud, adopting an average radial velocity of 40 km\,s$^{-1}$, and assuming an inclination angle $\theta$ from the plane of sky, proper motion due to the transverse velocity of the jet is then $\sim \frac{0.36}{\tan\theta}$\arcsec. The actual inclination angle of the system is not well determined but is estimated to be ranging from 15\arcdeg\ to 75\arcdeg\ and possibly $\sim45\arcdeg$ \citep{santiago2009}. Our low-velocity shell detections further imply a tighter range of 23\arcdeg\ to 67\arcdeg\ (Section \ref{subsec_LVShell}). The estimated proper motion is $\sim0\farcs36$ and should be no larger than $\sim0\farcs85$, which cannot fully account for the observed $\sim1.4\arcsec$ offset in the red-shifted jet. The jet velocity is quite symmetrical so that similar degrees of proper motion would be expected for both the red- and blue-shifted lobes. This is inconsistent with the current observation that the systematic offsets are only significant in the red-shifted lobe but not in the blue-shifted one.

As the single-dish-combined CO $J$ = 2--1 data of \citet{santiago2009} are compared to the interferometry-only CO $J$ = 3--2 data, one could ask whether the observed systematic offsets results from the missing flux in CO $J$ = 3--2. The answer may depend on the amount of missing flux, and whether the peak positions will be affected by the missing extended component or not. 
Unfortunately, due to the poor signal-to-noise ratio of the single dish JCMT data, we can not directly estimate the missing flux in the EHV regime. Therefore 
as a simple check, we examine the peak positions of CO $J$ = 2--1 emission in the PdBI-only maps and the single-dish-combined maps and confirm that the peak positions are not affected by the extended emission. This does not guarantee that the same is true for CO $J$ = 3--2 emission, but it is in favor of our argument. Through these considerations, we conclude that the offsets in the peak positions are likely real, although higher angular resolution CO $J$ = 2--1 data are still needed to resolve.

\subsection{I04166, L1448C, and HH 211 Outflows}% old 4.2

Now that I04166 has been studied by the SMA at a higher angular resolution, we can compare it with those well-studied Class 0 outflows such as HH 211 and L1448C. HH 211 is driven by a $\sim3.6$ $L_{\sun}$ Class 0 protostar in the IC 348 complex of Perseus ($d \sim280$ pc). It lies close to the plane of sky ($<10\arcdeg$) and has been one of the best candidates to search for jet rotation \citep[][and references therein]{lee2009}. L1448C, also located in the Perseus molecular cloud ($d \sim 250$ pc), is driven by a 7.5 $L_{\sun}$ Class 0 source and is thought to have an inclination angle of $\sim21\arcdeg$ from the plane of sky \citep[][and references therein]{hirano2010}. 
These outflows present common features of a less-collimated conical (V-shaped) shell at low velocities and a well-collimated jet-like component at higher velocities. Such characteristic appears to be the basic picture of an outflow system at its early stage. 

Aside from the similarity, the lower bolometric luminosity (0.4 $L_{\sun}$) and smaller envelope mass \citep{bontemps1996} suggest that I04166 may be on the lower-mass end among the class of the youngest objects. 
Some differences between I04166 and the others might be due to the amount of material (mass loss) contained in the system, or due to differences in excitation conditions or evolution phase. For example, the non-detection of SiO $J$ = 8--7 in our observation is in clear contrast to the well-mapped SiO $J$ = 8--7 emission in L1448C \citep{hirano2010}, and also to the detection of even higher SiO transition in HH 211 \citep{nisini2002}. The column densities of several other molecular species were also found to be systematically lower in I04166 than in L1448C \citep{tafalla2010}.
The CO observations also provide some hints. While the EHV jet and low velocity shell components of I04166 were found to be spectrally and morphologically distinct, substantial CO emission was detected between the two components in L1448C. This intermediate gas component appears like a transition between the wider V-shaped structure and the collimated jet \citep[][Figure 4]{hirano2010}, and could represent material of the underlying wide-angle wind responsible for producing the low velocity shell. The counterpart in I04166 may be invisible only because of  differences in excitation conditions. We finally note that given a mean radial velocity of 40 km\,s$^{-1}$ and an inclination angle of 45\arcdeg, the de-projected jet velocity of $\sim55$ km\,s$^{-1}$ in I04166 also appears lower than the typical jet velocity of 100--200 km\,s$^{-1}$. The mean de-projected jet velocity of HH 211, for example, was $\sim170$ km\,s$^{-1}$ \citep{lee2009}. 
  
\subsubsection{Characteristic Pattern and Head--Tail Structure of the EHV Knots} %old 4.2.1   
    
The velocity gradients along the outflow axis within the EHV knots form the characteristic pattern in the PV diagram of the I04166 outflow.  Similar velocity gradients can also be found in HH 211 and L1448C if one carefully inspects the data. SiO $J$ = 8--7 and CO $J$ = 3--2 observations with the SMA have revealed similar features at positions of  BI-c, BII-a, RI-b, RI-c, and RII-a in L1448C \citep[][Figure 7]{hirano2010}, and at positions of BK2 and BK3 in HH211 \citep[][Figure 7]{lee2009}. Although bearing some resemblance, their overall kinematic patterns are much more irregular and complicated than that of I04166.  A few pulsed-jet models \citep[e.g., ][]{raga1990,stone1993} have been invoked to explain the appearance of a sawtooth pattern, however, the unique time sequence and standing structure embedded in the pattern of I04166 may hint some different mechanisms from a simple scenario.

Another common feature found in both I04166 and HH 211 is the ``broad head--narrow tail'' structure. These are also visible in knots BK2, BK3, and RK3 of HH211. One interpretation may be that the head consists of material ejected sideways at the internal shocks and the tail represents the weakly shocked material in the jet beam \citep{lee2009}.  The broad-head part of a knot will consist of the sideways squirted gas in this context. The lateral velocity component (directed perpendicular to the jet axis) was proposed to explain the sawtooth pattern in \citet{santiago2009}. However, as our CO $J$ = 3--2 PV diagram shows, the gradient feature extends up to the narrow high-velocity tail of EHV knots where sideway velocity is thought to be insignificant or at least not yet well developed. Therefore in explaining the characteristic velocity gradients, the variational velocities in the axial direction of the weakly-shocked gas may also play a role.

\section{SUMMARY AND CONCLUSIONS}\label{sec_conclusion}

\begin{figure}[ht]
        %\figurenum{}
        \includegraphics[width=0.9\textwidth]{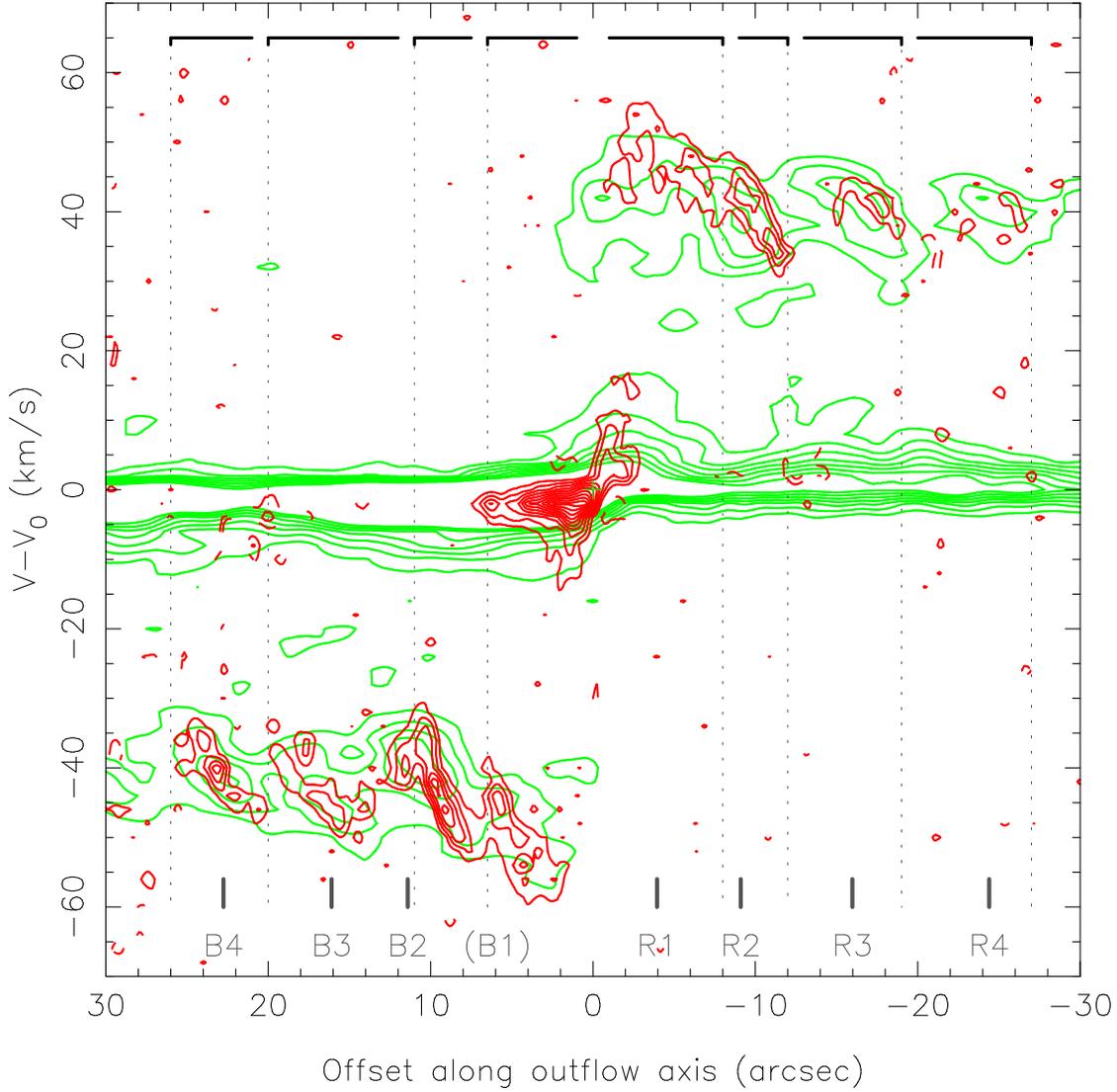}
        \caption{Position--velocity diagram of CO $J$ = 3--2 emission (red contours) along the outflow axis overlaid on CO $J$ = 2--1 (gray contours). For CO $J$ = 3--2, contour levels are from 3$\sigma$ in steps of 3$\sigma$, where 1$\sigma$ is 37 mJy per $1\farcs02\times0\farcs80$ beam.  For CO $J$ = 2--1, contour levels are also from 3$\sigma$ in steps of 3$\sigma$,  where 1$\sigma$ is 50 mJy per 3\arcsec beam. The short vertical line segments show positions of knots previously identified in CO $J$ = 2--1 observations \citep[Table 1 of ][]{santiago2009}, while the vertical dotted lines show the divided knot sections based on kinematics.
        The velocity structure of the two innermost EHV knots (B1 and R1) is clearly seen, and knot B1 can now be clearly identified from B2 under in the high resolution data. Note that within the EHV regime CO $J$ = 3--2 emission tends to concentrate toward the ``head'' of each knot compared to CO $J$ = 2--1 in the red-shifted lobe.   
        }
        \label{fig_co32co21_pv}
\end{figure}

\begin{figure}[ht]
        %\figurenum{}
        \includegraphics[width=0.65\textwidth]{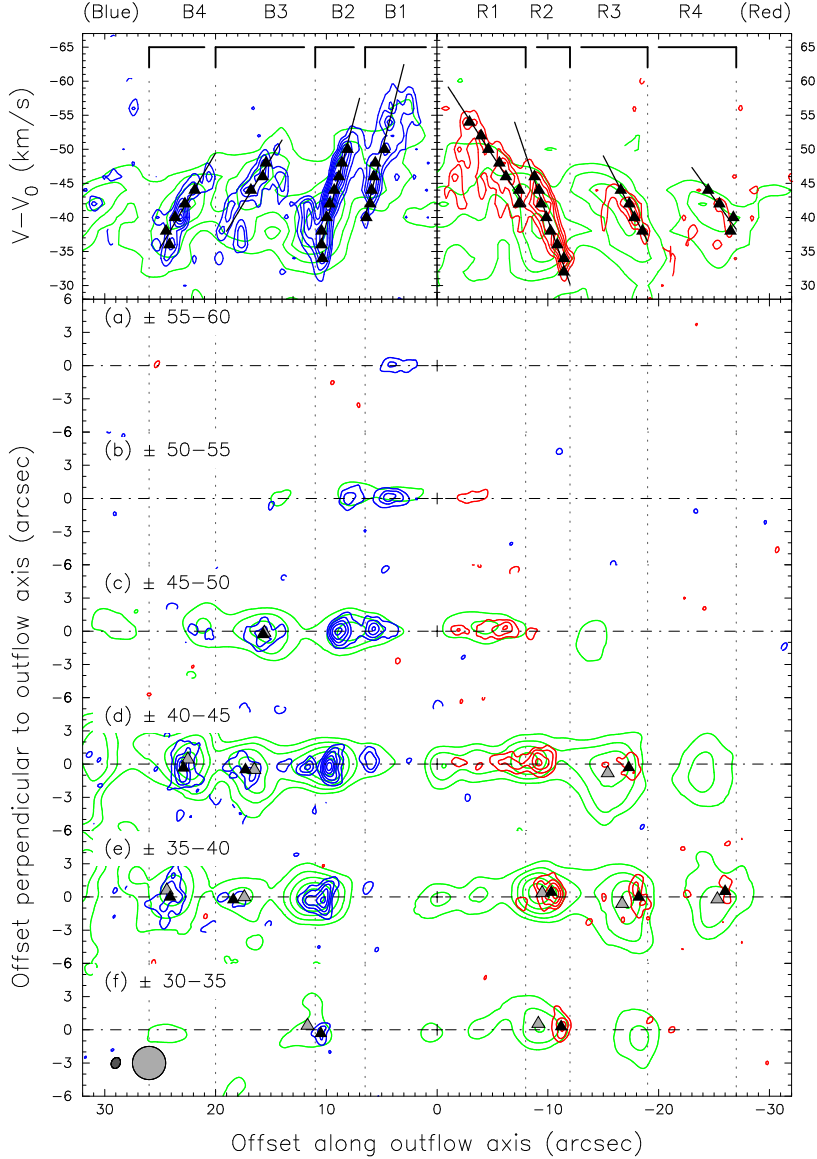}
        \caption{\textit{(a)--(f)}: Channel maps of CO $J$ = 3--2 emission (blue/red contours) are overlaid on top of CO $J$ = 2--1 (green contours) with 5 km\,s$^{-1}$ velocity bin. The velocities of each channel with respect to systemic velocity are shown on the upper left corner. For CO $J$ = 3--2, contour levels are from 4$\sigma$ in steps of 3$\sigma$, where 1$\sigma$ is 23 mJy per beam. For CO $J$ = 2--1, contour levels are from 5$\sigma$ in steps of 3$\sigma$, where 1$\sigma$ is 32 mJy per beam. The beam sizes are shown on the bottom left corner of the whole plot. The dark and gray solid triangles on the maps mark the fitted knot positions of CO $J$ = 3--2 and CO $J$ = 2--1 emission respectively.         
         \textit{Top}: PV diagrams of EHV range is shown. For CO $J$ = 3--2, contour levels are from 3$\sigma$ in steps of 2$\sigma$, where 1$\sigma$ is 37 mJy per beam. For CO $J$ = 2--1, contour levels are from 3$\sigma$ in steps of 3$\sigma$, where 1$\sigma$ is 50 mJy per beam. Solid triangles mark the fitted positions and the solid dark lines are the linear fits to them.
        The vertical dotted lines and the horizontal bars at the top show the division of knot domains based on the velocity pattern (see text). Note the slow-broad head and fast-narrow tail structure within knots B2 and R2. Also note that for R2 and R3 in panels (d)--(f), CO $J$ = 3--2 emission appears to concentrate more toward the head of the knots compared to CO $J$ = 2--1.
         }
        \label{fig_co32co21_pv_and_ehv_5kms_chmap_hori}
\end{figure}

\begin{figure}[ht]
        \includegraphics[width=\textwidth]{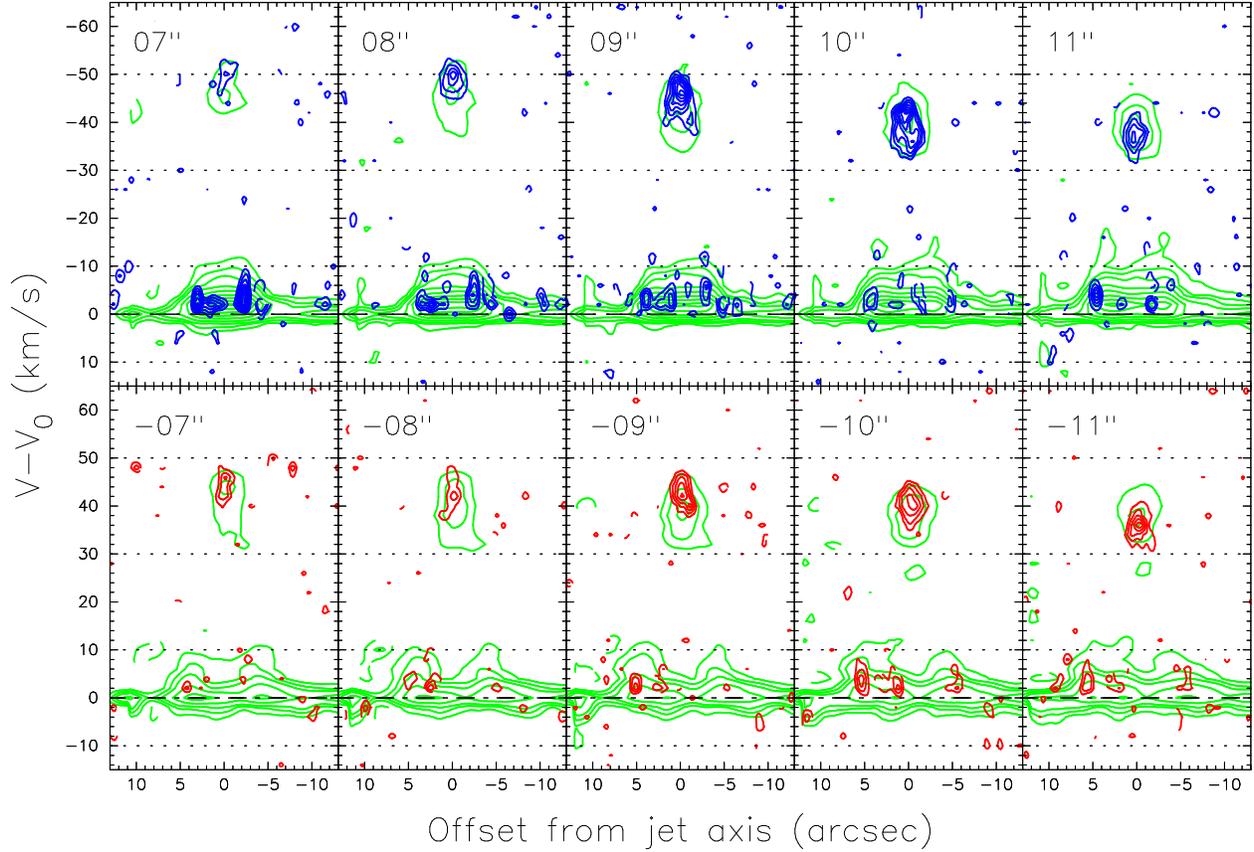}
        \caption{Position--Velocity cuts perpendicular to the outflow axis at various positions along outflow axis. The position offsets from the center are shown in the upper left corner of each panel. For CO $J$ = 3--2 (blue/red contours), contour levels are from 3$\sigma$ in steps of 2$\sigma$, where 1$\sigma$ is 37 mJy per beam. For CO $J$ = 2--1 (green contours), contour levels are drawn at 4$\sigma$, 8$\sigma$, 12$\sigma$, 20$\sigma$, 30$\sigma$,\ldots, where 1$\sigma$ is 50 mJy per beam.
        }
        \label{fig_co32co21_pv_crosscut}
\end{figure}

\begin{figure}[ht]
        \includegraphics[width=0.75\textwidth]{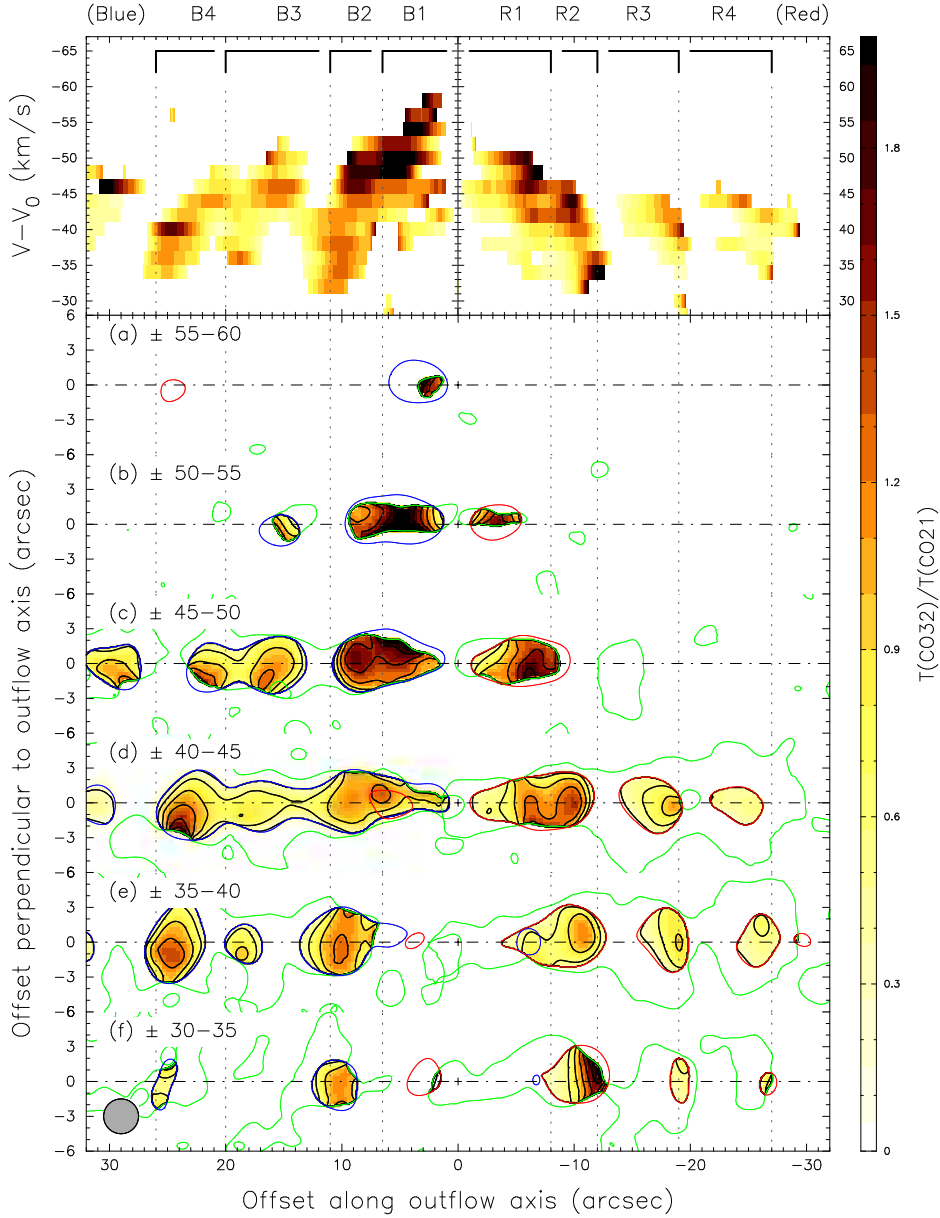}
        \caption{\textit{(a)--(f)}: CO $J$ = 3--2 / CO $J$ = 2--1 line ratio maps (linear color scale and dark contours) are shown with 5 km\,s$^{-1}$ velocity bins. Contour levels are from 0.3 in steps of 0.3. The CO $J$ = 3--2 data were smoothed to the same resolution as the CO $J$ = 2--1 data (3\arcsec\ FWHM) before taking the line ratio. The red/blue contour lines show the 3$\sigma$ level of CO $J$ = 3--2 data, and the green contour lines that of CO $J$ = 2--1 data. Line ratios are only calculated at positions where both emission lines are brighter than their respective 3-sigma levels.
        \textit{Top}: Line ratios in PV diagram made from data of 2 km\,s$^{-1}$ velocity bin are plotted. The labelling is similar to that in figure \ref{fig_co32co21_pv_and_ehv_5kms_chmap_hori}. Note that higher line ratios are found in knots B1 and R1 and at higher velocity in panels (a)--(c). A gradient is found most apparent within R2 and R3 in panels (d)--(f).
                 }
        \label{fig_co32co21_pv_and_ehv_5kms_chmap_hori_ratio}
\end{figure}

\begin{figure}[ht]
        \includegraphics[width=0.8\textwidth]{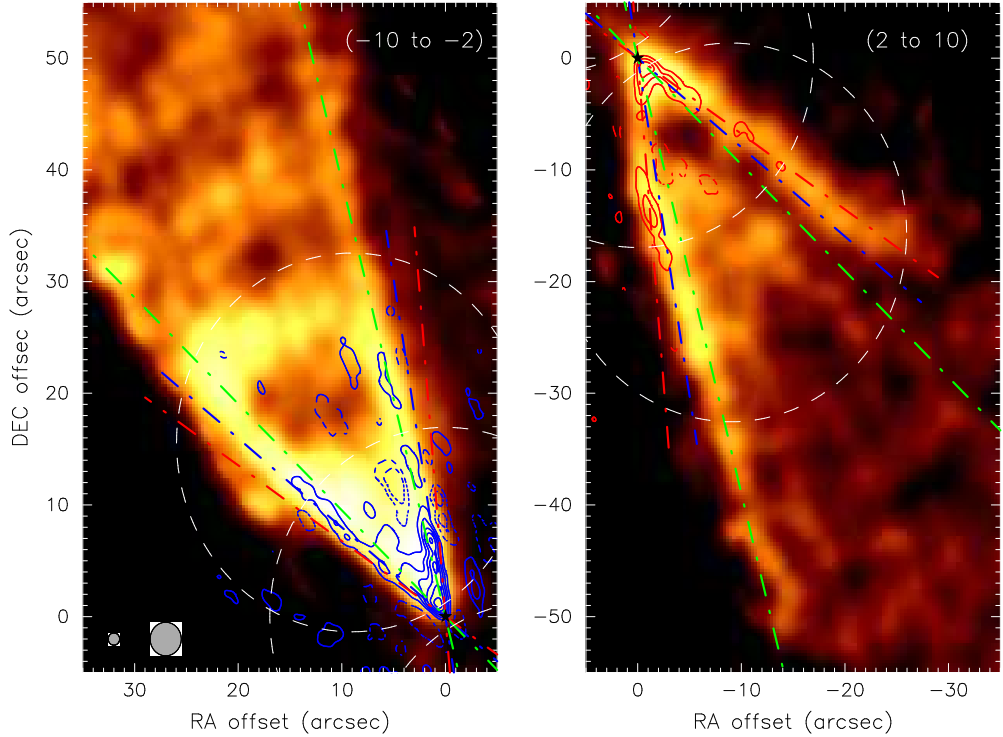}
        \caption{Integrated CO $J$ = 3--2  emission (blue/red contours) of I04166 outflow overlaid on that of CO $J$ = 2--1 emission (shown in color scale) at the low velocity regime (2$<|$V$-$V$_0|<10$ km\,s$^{-1}$). The CO $J$ = 3--2 contours are the same as those in the left panel of Figure \ref{fig_co32_mom0_SHVEHV}, except that the blue- and red-shifted lobes are shown separately here. The opening angle of the V-shape structure obtained for the blue- and red-shifted CO $J$ = 3--2 maps are illustrated with dot-dashed lines of the corresponding color, i.e., 42{\arcdeg} for blue-shifted and  51{\arcdeg} for red-shifted lobes. The green dot-dashed lines show the 32{\arcdeg} opening angle previously determined by \citet{santiago2009}. The white dashed circles show the primary beam HPBW of our observations. The beamsize of CO $J$= 3--2 and CO $J$= 2--1 data are shown at the lower-left corner of the left panel.
        }
        \label{fig_co32co21_mom0_compare_SHV_opening}
\end{figure}

\begin{figure}[ht]
        \includegraphics[width=0.5\textwidth]{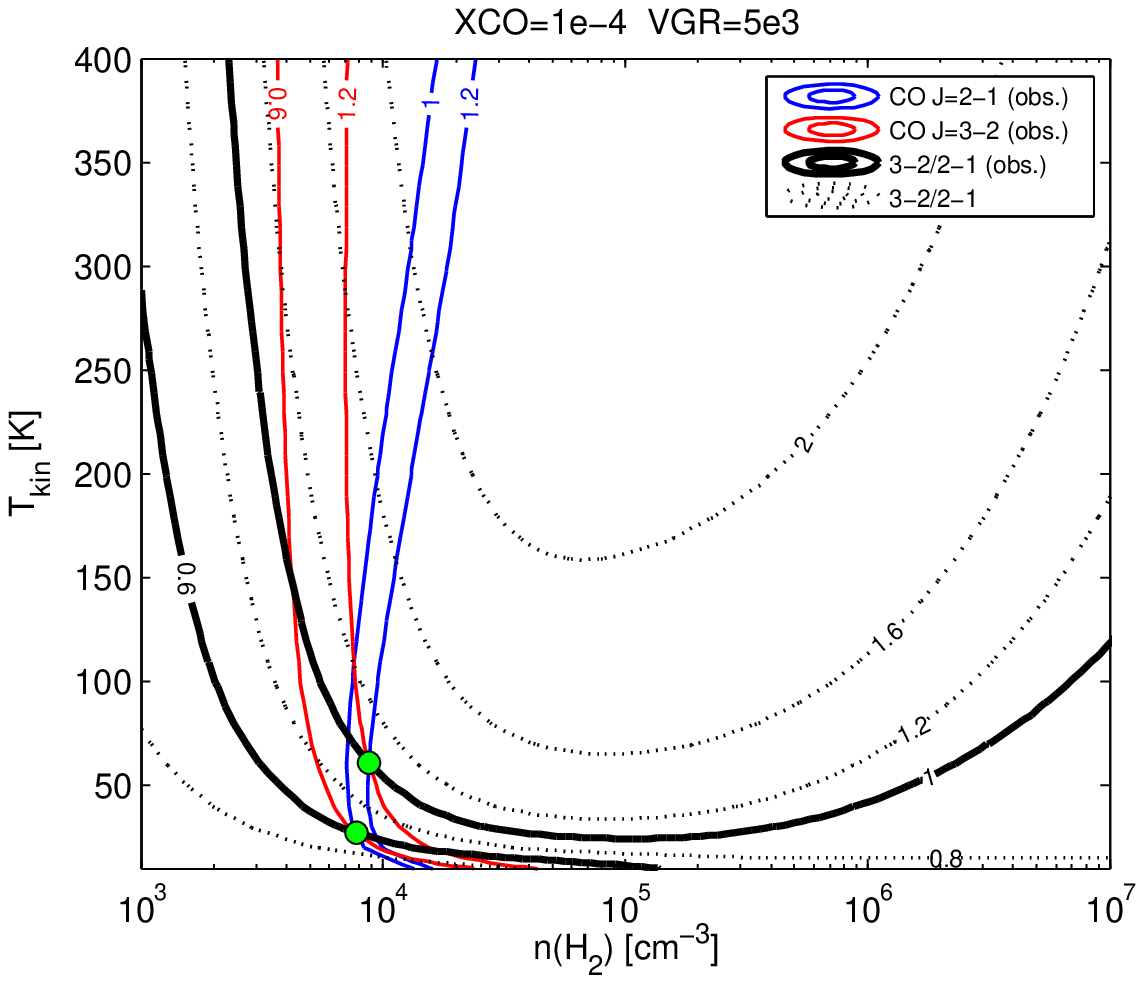}
        \includegraphics[width=0.5\textwidth]{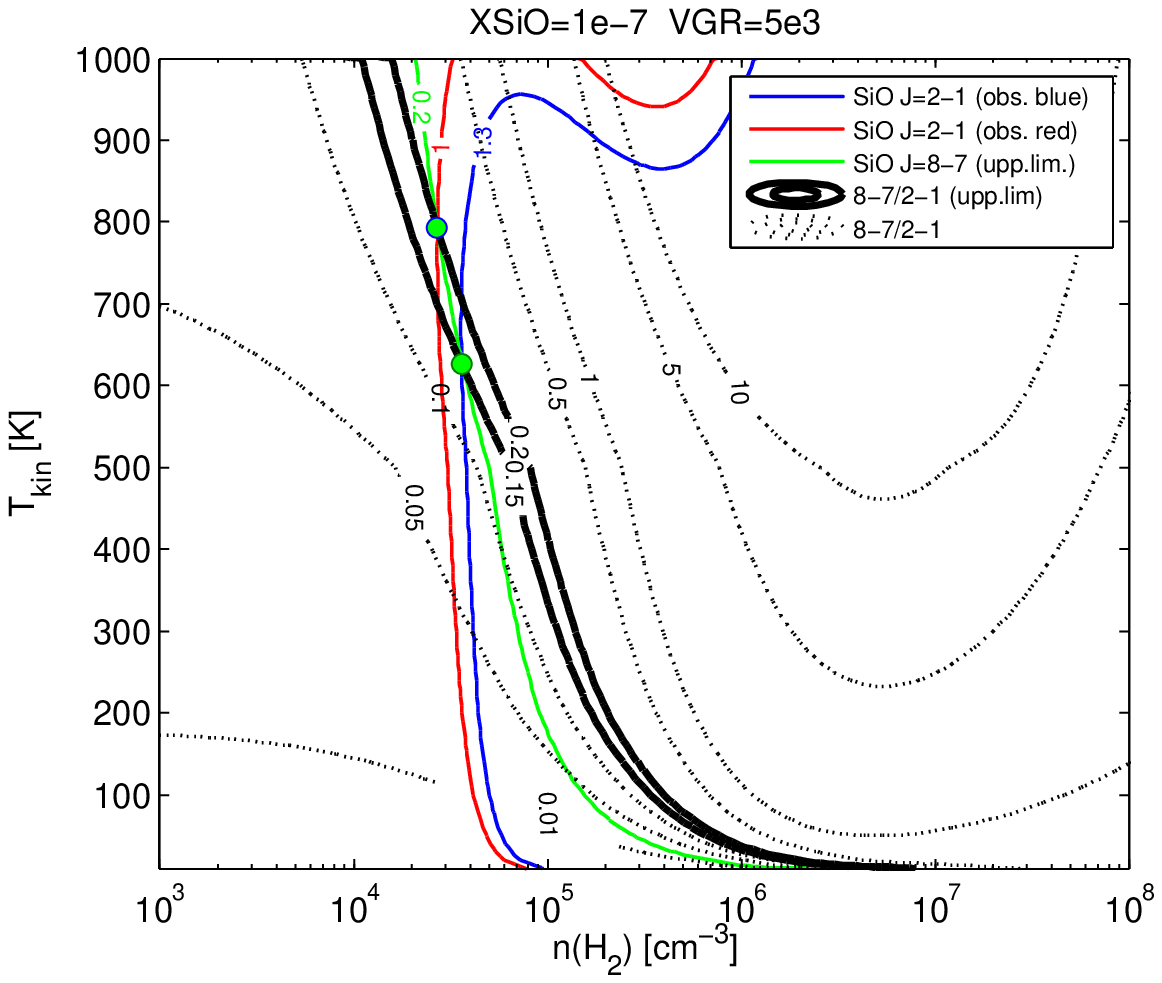}
        \caption{Large velocity gradient (LVG) analysis of CO (\textit{left}) and SiO emission (\textit{right}). In the left panel, the dark dotted contours show the general trend of CO $J$ = 3--2/$J$ = 2--1 line ratio. The contour levels start from 0.4 in increment of 0.4. The thick solid lines show some typical flux values from knots R2 and R3 we adopted to examine the possible head--tail differences in physical condition of a knot. The corresponding brightness temperatures of CO $J$ = 2--1 and CO $J$ = 3--2  are also shown with red and blue solid lines, respectively.  The figure suggest a possible difference in gas kinetic temperature between head and tail. In the right panel, the dark dotted contours show the general trend of SiO $J$ = 8--7/$J$ = 2--1 line ratio. Contour levels are 0.01, 0.05, 0.1, 0.5, 1, 5, and 10. The thick solid lines show the observed upper limit of line ratio values set by the SiO $J$ = 2--1 peak emissions (blue/red solid lines) and the non-detection of SiO $J$ = 8-7 in this work (the green solid line). The results suggest an upper limit of gas kinetic temperature of $\sim700$ K. The filled green circles mark the constrained temperature-- density condition according to the observations. More details are in section \ref{subsec_lvg_ehv}.    
                 }
        \label{fig_i04166_lvg}
\end{figure}

\begin{figure}[ht]
        \includegraphics[width=0.5\textwidth]{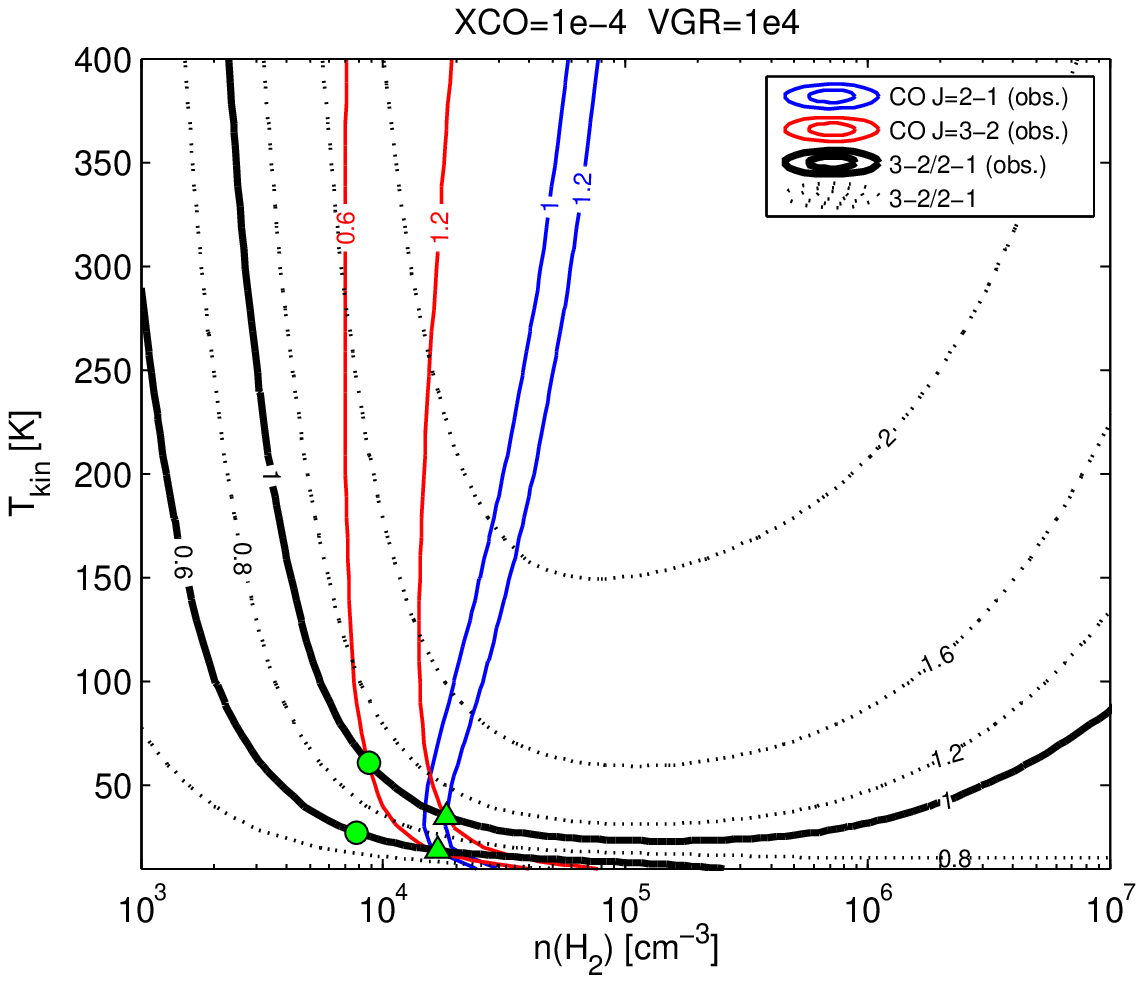}
        \includegraphics[width=0.5\textwidth]{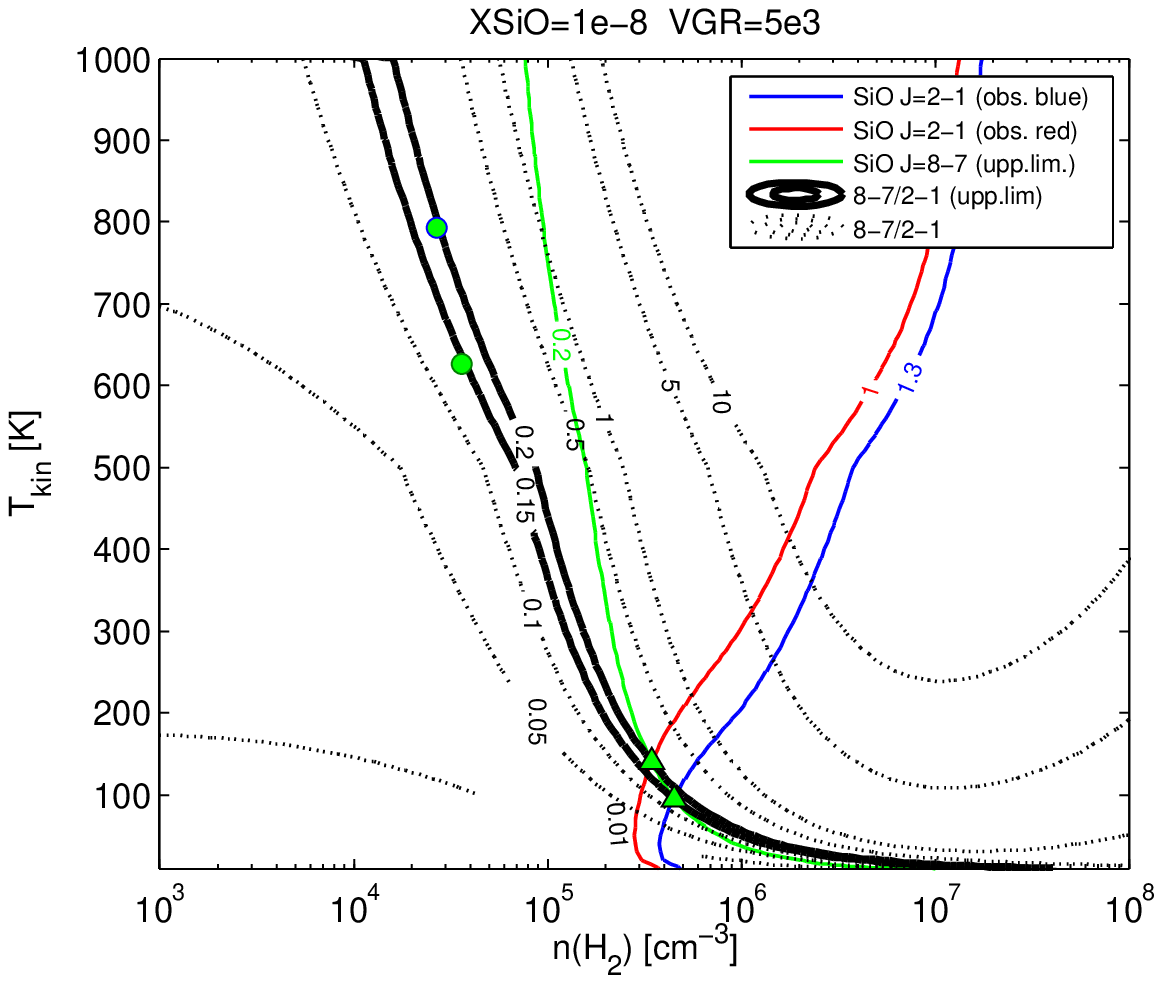}
        \includegraphics[width=0.5\textwidth]{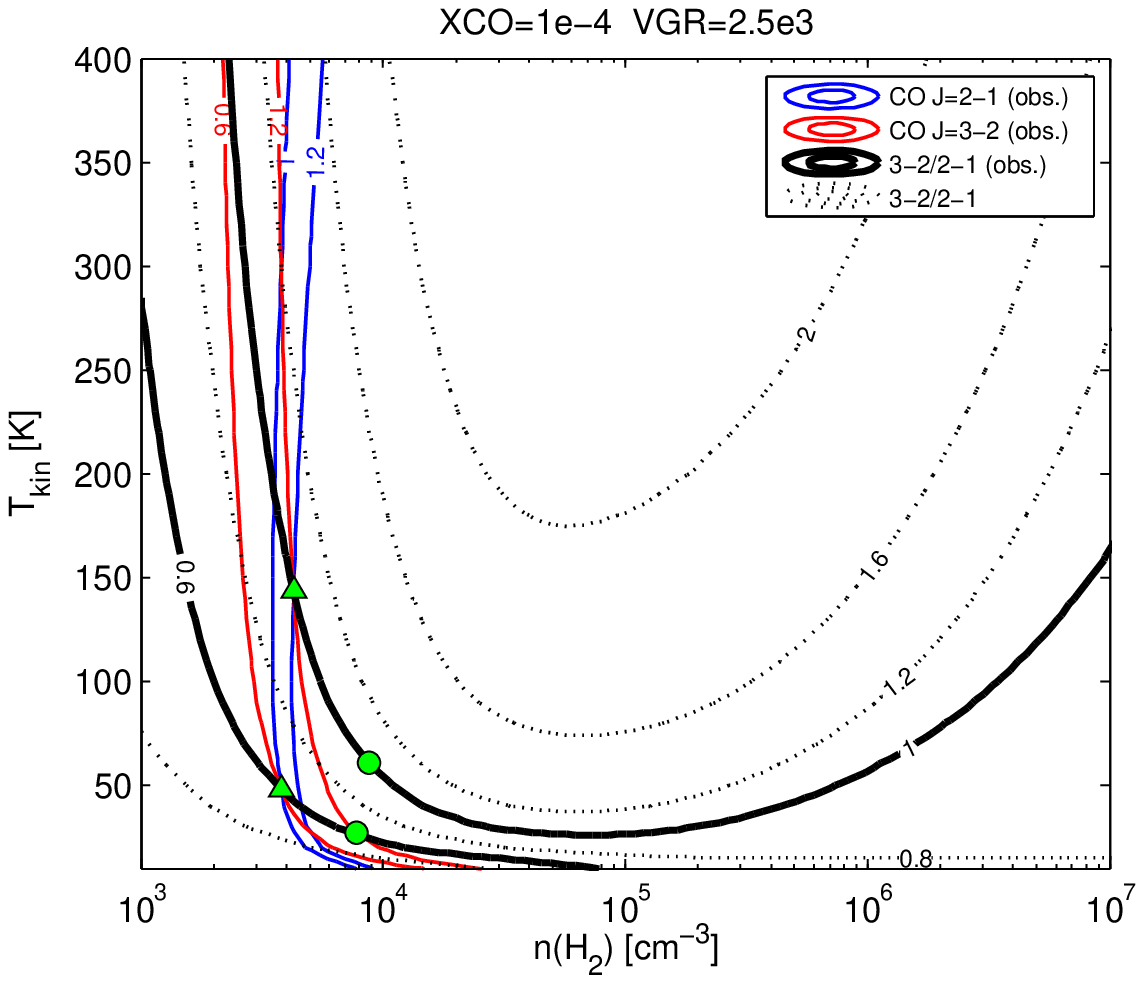}
        \includegraphics[width=0.5\textwidth]{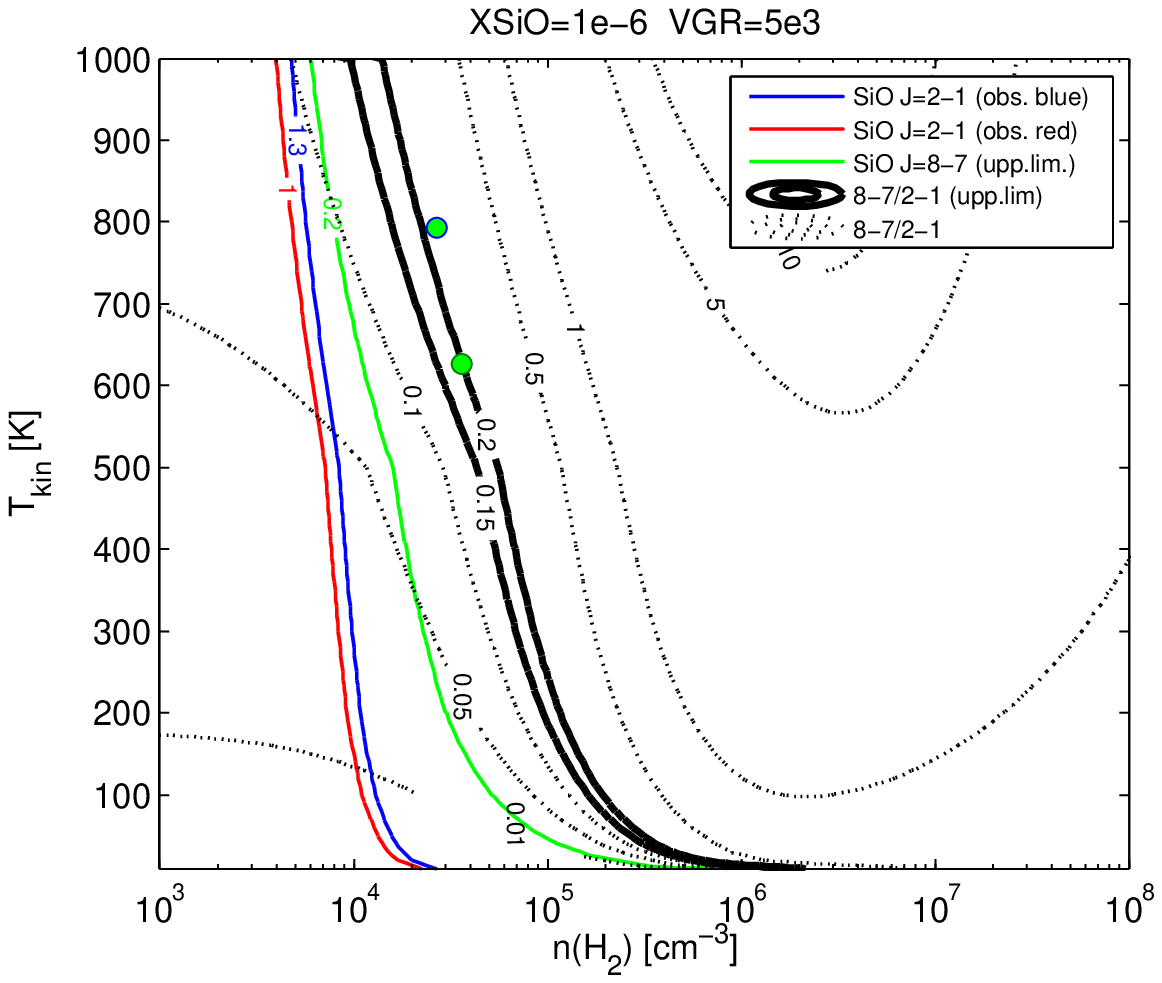}        
        \caption{Large velocity gradient (LVG) analysis of CO (\textit{left column}) and SiO emission (\textit{right column}) as in Figure \ref{fig_i04166_lvg} but with varied input LVG parameters. The velocity gradient (VGR) is varied up and down by a factor of two in the CO calculation and the fractional abundance by an order of magnitude in the SiO calculation. The adopted parameters are labelled on top of each figure. The contours have the same meaning as in Figure \ref{fig_i04166_lvg}, but here we mark the constrained results with green triangles instead of green circles, which denote the constrained values in the reference cases (in Figure \ref{fig_i04166_lvg}) for comparison.   
                 }
        \label{fig_i04166_lvg_uncertainty}
\end{figure}

        We have presented interferometric observations toward the I04166 molecular outflow at a $\sim1\arcsec$ resolution with the SMA. We mapped the system in the 350 GHz continuum and CO $J$ = 3--2 emission. No SiO $J$ = 8--7 emission was detected in our observations. Using also CO $J$ = 2--1 data from previous studies, we examine the kinematic features of the EHV jet and the low velocity shell of I04166, and discuss some of the properties in common with other Class 0 outflows such as HH 211 and L1448C. Our main conclusions are the following:
        
\begin{enumerate}
\item A single peak was detected in the 350 GHz continuum map. The visibility data show that there are at least two components with similar flux, but with sizes that differ by an order of magnitude. Such feature is also found in other Class 0 sources.
\item In the CO $J$ = 3--2 channel maps, a V-shaped conical structure is present in the low-velocity range ($|V-V_0|<10$ km\,s$^{-1}$), while highly-collimated knots are detected in the EHV range ($|V-V_0|>30$ km\,s$^{-1}$). This is in general agreement with previous CO $J$ = 2--1 data.
\item Thanks to our higher-angular-resolution observations, we identified the most recent blue-shifted knot B1 $\sim6\arcsec$ to the northeast of the central source, completing the symmetry of the jet as the counterpart of knot R1. The maximum velocities of the two youngest knots (B1 and R1) are found to be the highest among the series.

\item The PV diagram along the outflow axis reveals the same ``slow head--fast tail'' sawtooth velocity pattern of EHV knots observed in CO $J$ = 2--1. This is roughly consistent with an interpretation based on a pulsed jet scenario. The CO $J$ = 3--2 data further suggest the variational axial velocities could be directly involved.

\item The CO $J$ = 3--2 channel maps reveal a ``broad head--narrow tail'' structure of the EHV knots, where the slower gas at the head is more laterally elongated than the faster gas at the tail. Similar structures were found in HH 211.  

\item PV cuts perpendicular to the outflow axis indicate that the opening angle of the V-shell remains almost the same over the lowest velocities from $|V-V_0|= 0$ to 8 km\,s$^{-1}$. This could be understood if the emission represents a shear flow of accelerated ambient gas moving along the wall of evacuated cavities on both sides, also consistent with the previous CO $J$ = 2--1 findings.

\item The average opening angle of the CO $J$ = 3--2 blue-shifted (red-shifted) V-shaped structure is found to be $\sim42\arcdeg$ ($\sim51\arcdeg$), which appears greater than the 32\arcdeg\ determined with CO $J$ = 2--1. Direct comparisons between the two maps (Figure \ref{fig_co32co21_mom0_compare_SHV_opening}) show that this may be due to the smaller coverage of the CO $J$ = 3--2 data, which trace only the shell structure closer to the protostar. The shell is more parabolic at distances beyond the V-shaped base.  

\item In the 5 km\,s$^{-1}$ velocity binned maps (Figure \ref{fig_co32co21_pv_and_ehv_5kms_chmap_hori}), comparisons of the peak positions show that on average the red-shifted EHV CO $J$ = 3--2 emission peaks $\sim1\farcs4$ farther from the central source than the CO $J$ = 2--1 emission, and the offsets are unlikely due to proper motions alone. An alternative view of this phenomenon is that CO $J$ = 3--2 emission appears systematically brighter at higher velocities at a given position (Figure \ref{fig_co32co21_pv}), which might suggest a difference in the excitation of the emitting gas.

\item Maps of CO $J$ = 3--2/CO $J$ = 2--1 line ratio (Figure \ref{fig_co32co21_pv_and_ehv_5kms_chmap_hori_ratio}) reveal higher line ratios at the inner knots and at higher speeds. An internal gradient with higher line ratios found toward the head of each knot is present in the red-shifted lobe. Our simple LVG analysis suggests a possible temperature structure within each knot where the head has a higher gas kinetic temperature.

\item SiO $J$ = 8--7 emission was not detected in our SMA observations. Using the upper limit imposed by the SiO $J$ = 8--7 and the PdBI SiO $J$ = 2--1 data, a LVG analysis assuming fractional abundance of 10$^{-7}$ suggests a kinetic temperature of $<700$ K.

\end{enumerate}

Finally, we note that its proximity (Taurus molecular cloud) and relatively clean environment makes I04166 one of the best targets to study outflow kinematics and knot propagation. Further observations such as higher rotational transition lines of CO at a comparable or higher spatial resolution will help constrain the physical conditions of the outflow and resolve some uncertainties. SiO mapping toward the inner part of the jet at high angular resolution will help understand the thermal-chemical history and excitation conditions of the outflow.   

\acknowledgments

This work was supported by funds from the Theoretical Institute for Advanced Research in Astrophysics (TIARA) through the Academia Sinica and student stipends from the National Science Council of Taiwan.

%%%%%%%%%%%%%
% REFERENCE %
%%%%%%%%%%%%%

%%%%%%%%%%%%
%  Tables  %
%%%%%%%%%%%%
\clearpage

%%%% Table 1
%%%% beam sizes and rms noise levels
\begin{deluxetable}{ccccccc}
        \tablewidth{0pt}% natural width
        \tablecaption{Beam sizes and rms noise levels of the data.\label{tab_bsize_rms}}
        \tablehead{
                \colhead{Line} &\colhead{Bmaj} &\colhead{Bmin} &\colhead{P.A.} &\colhead{1$\sigma$} &\colhead{Channel width}  \\
                \colhead{} &\colhead{(\arcsec)} &\colhead{(\arcsec)} &\colhead{(\arcdeg)} &\colhead{(mJy\,beam$^{-1}$)} &\colhead{(km\,s$^{-1}$)} 
        }
\startdata
350 GHz Continuum & 0.85 & 0.59 & $-85.1$ & 2.8 & \nodata \\
CO $J$ = 3--2 & 1.02 & 0.8 & $-84.9$ & 37 & 2  \\
%HCO$^+$ $J$ = 4--3 & 0.99 & 0.81 & $-89.5$ & 89 & 0.5 \\
\enddata
 
\end{deluxetable}

%%%%% Table 2
%%%%% Fitted slope
%\begin{deluxetable}{ccc}
%        \tablewidth{0pt}% natural width
%        \tablecaption{Fitted Slopes of  the Sawtooth Structure.\label{tab_slope}}
%        \tablehead{
%                \colhead{Knot} &\colhead{Fitted Slope} &\colhead{Fitted Slope}  \\
%                \colhead{(\#)} &\colhead{(km\,s$^{-1}$\,arcsec$^{-1}$)} & \colhead{(10$^3$ km\,s$^{-1}$\,pc$^{-1}$)}
%        }
%\startdata                  % slope  sig
%B1 & $6.4\pm1.2$ & $9.4\pm1.8$ \\  % 6.387+-1.17
%B2 & $6.1\pm0.6$ & $8.9\pm0.9$ \\  % 6.130+-0.58
%B3 & $2.7\pm0.9$ & $4.0\pm1.3$ \\  % 2.677+-0.86
%B4 & $2.8\pm0.6$ & $4.1\pm0.9$ \\  % 2.786+-0.61
%\tableline 
%R1 & $2.5\pm0.2$ & $3.7\pm0.3$ \\  % 2.505+-0.17
%R2 & $4.8\pm0.3$ & $7.1\pm0.4$ \\  % 4.758+-0.29
%R3 & $3.1\pm0.2$ & $4.6\pm0.3$ \\  % 3.133+-0.20
%R4 & $2.3\pm0.7$ & $3.4\pm1.0$     % 2.263+-0.73
%\enddata
%\end{deluxetable}

%%%%%%%%%%%%%
%  Figures  %
%%%%%%%%%%%%%
\clearpage
%%%% Figure 1 
%%%% continuum map
        
%%%% Figure 2
%%%% continuum visibility fitting: two circular Gaussian

%%%% Figure 3 (modified)
%%%% CO J=3-2 6 kms channle maps

%%%% Figure 4 (new/modified)
%%%% CO32 SHVEHV map

%%%% Figure 5 (corrected)
%%%% Missing flux CO J=3-2 spectrum at 3 position (with single dish data)

%%%% Figure 6 (modified)
%%%% CO J=3-2 CO J=2-1 PV diagram

%%%% Figure 7 (slightly modified)
%%%% CO J=3-2 and CO J=2-1 EHV 5kms channle map horizontal with PV

%%%% Figure 8 (prev. Figure 10) 
%%%% PV crosscut co32co21

%%%% Figure 9  (slightly modified, prev. Figure 11) 
%%%% CO J=3-2conv3/CO J=2-1 EHV 5kms ratio map horizontal with PV (line ratio)

%%%% Figure 10 (New)
%%%% Opening angle of Shell
%%%% CO32  CO21 SHV maps and opening angle

%%%% Figure 11 (New)
%%%% LVG
%%%% CO J=3-2conv3/CO J=2-1 
%%%% SiO J=8-7(upper lim)/SiO J=2-1

%%%% Fig 12 (New)
%%%% LVG-2 
%%%% CO J=3-2conv3/CO J=2-1 
%%%% SiO J=8-7(upper lim)/SiO J=2-1


\begin{thebibliography}{}
\bibitem[Beckwith et al.(1990)]{beckwith1990} Beckwith, S.~V.~W., Sargent, A.~I., Chini, R.~S., \& Guesten, R.\ 1990, \aj, 99, 924 % dust opacity power-law
\bibitem[Bontemps et al.(1996)]{bontemps1996} Bontemps, S., Andre, P., Terebey, S., \& Cabrit, S.\ 1996, \aap, 311, 858
\bibitem[Codella et al.(2007)]{codella2007} Codella, C., Cabrit, S., Gueth, F., et al.\ 2007, \aap, 462, L53  % HH212
\bibitem[Davis et al.(2010)]{davis2010} Davis, C.~J., Chrysostomou, A., Hatchell, J., et al.\ 2010, \mnras, 405, 759 % I04166 JCMT
\bibitem[Dayou \& Balan{\c c}a(2006)]{dayou2006} Dayou, F., \& Balan{\c c}a, C.\ 2006, \aap, 459, 297 % LVG SiO coll. data
\bibitem[Dent et al.(1998)]{dent1998} Dent, W.~R.~F., Matthews, H.~E., \& Ward-Thompson, D.\ 1998, \mnras, 301, 1049 % Dust opacity observation
\bibitem[Flower(2001)]{flower2001} Flower, D.~R.\ 2001, Journal of Physics B Atomic Molecular Physics, 34, 2731 % LVG CO coll. data
\bibitem[Goldreich \& Kwan(1974)]{goldreich1974} Goldreich, P., \& Kwan, J.\ 1974, \apj, 189, 441 %LVG
\bibitem[Ho et al.(2004)]{ho2004} Ho, P.~T.~P., Moran, J.~M., \& Lo, K.~Y.\ 2004, \apjl, 616, L1 %SMA
\bibitem[Hirano et al.(2006)]{hirano2006}
Hirano, N., Liu, S.-Y., Shang, H., Ho, P.~T.~P., Huang, H.-C., Kuan, Y.-J., McCaughrean, M.~J., \& Zhang, Q.\ 2006, \apjl, 636, L141
\bibitem[Hirano et al.(2010)]{hirano2010} Hirano, N., Ho, P.~P.~T., Liu, S.-Y., et al.\ 2010, \apj, 717, 58 % L1448C
\bibitem[Konigl \& Pudritz(2000)]{konigl2000} Konigl, A., \& Pudritz, R.~E.\ 2000, Protostars and Planets IV, 759 % Disk-wind
\bibitem[Lee et al.(2007a)]{lee2007a} Lee, C.-F., Ho, P.~T.~P., Hirano, N., et al.\ 2007, \apj, 659, 499 % HH212
\bibitem[Lee et al.(2007b)]{lee2007b} Lee, C.-F., Ho, P.~T.~P., Palau, A., et al.\ 2007, \apj, 670, 1188 % HH211
\bibitem[Lee et al.(2009)]{lee2009} Lee, C.-F., Hirano, N., Palau, A., et al.\ 2009, \apj, 699, 1584 % HH211
\bibitem[Meyers-Rice \& Lada(1991)]{meyers1991} Meyers-Rice, B.~A., \& Lada, C.~J.\ 1991, \apj, 368, 445
\bibitem[Nisini et al.(2002)]{nisini2002} Nisini, B., Codella, C., Giannini, T., \& Richer, J.~S.\ 2002, \aap, 395, L25 % HH211high-J SiO
\bibitem[Nisini et al.(2007)]{nisini2007} Nisini, B., Codella, C., Giannini, T., et al.\ 2007, \aap, 462, 163 % LVG
%\bibitem[Oka et al.(1998)]{oka1998} Oka, T., Hasegawa, T., Hayashi, M., Handa, T., \& Sakamoto, S.\ 1998, \apj, 493, 730 % LVG
\bibitem[Palau et al.(2006)]{palau2006} Palau, A., et al.\ 2006, \apjl, 636, L137
\bibitem[Raga et al.(1990)]{raga1990} Raga, A. C., Binette, L., Canto, J., \& Calvet, N. 1990, \apj, 364, 601
\bibitem[Santiago-Garc{\'{\i}}a et al.(2009)]{santiago2009} Santiago-Garc{\'{\i}}a, J., Tafalla, M., Johnstone, D., \& Bachiller, R.\ 2009, \aap, 495, 169 % I04166 CO21
\bibitem[Sault et al.(1996)]{sault1996} Sault, R.~J., Staveley-Smith, L., \& Brouw, W.~N.\ 1996, \aaps, 120, 375 % MIRIAD
\bibitem[Sch{\"o}ier et al.(2004)]{schoeier2004} Sch{\"o}ier, F.~L., J{\o}rgensen, J.~K., van Dishoeck, E.~F., \& Blake, G.~A.\ 2004, \aap, 418, 185  % L1448C
\bibitem[Sch{\"o}ier et al.(2005)]{schoeier2005} Sch{\"o}ier, F.~L., van der Tak, F.~F.~S., van Dishoeck, E.~F., \& Black, J.~H.\ 2005, \aap, 432, 369 %LVG CO/SiO 
\bibitem[Scoville et al.(1993)]{scoville1993} Scoville, N.~Z., Carlstrom, J.~E., Chandler, C.~J., et al.\ 1993, \pasp, 105, 1482 % MIR        
\bibitem[Shang et al.(2006)]{shang2006} Shang, H., Allen, A., Li, Z.-Y., et al. 2006, \apj, 649, 845
\bibitem[Shu et al.(2000)]{shu2000} Shu, F.~H., Najita, J.~R., Shang, H., \& Li, Z.-Y.\ 2000, Protostars and Planets IV, 789 % X-wind
\bibitem[Stone \& Norman (1993)]{stone1993} Stone, J. M., \& Norman, M. L. 1993, \apj, 413, 210
\bibitem[Surdej(1977)]{surdej1977} Surdej, J.\ 1977, \aap, 60, 303 
\bibitem[Tafalla et al.(2004)]{tafalla2004} Tafalla, M., Santiago, J., Johnstone, D., \& Bachiller, R.\ 2004, \aap, 423, L21  % I04166 CO21
\bibitem[Tafalla et al.(2010)]{tafalla2010} Tafalla, M., Santiago-Garc{\'{\i}}a, J., Hacar, A., \& Bachiller, R.\ 2010, \aap, 522, A91 % I04166
\bibitem[Wernli et al.(2006)]{wernli2006} Wernli, M., Valiron, P., Faure, A., et al.\ 2006, \aap, 446, 367 %LVG CO coll. data




\end{thebibliography}
\end{document}